\documentclass[journal,comsoc]{IEEEtran}
\usepackage{amsmath,amsfonts}
\usepackage{algorithmic}
\usepackage{array}
\usepackage[caption=false,font=normalsize,labelfont=sf,textfont=sf]{subfig}
\usepackage{textcomp}
\usepackage{stfloats}
\usepackage{url}
\usepackage{bm}
\usepackage{multirow}
\usepackage{makecell}
\usepackage{verbatim}
\def\BibTeX{{\rm B\kern-.05em{\sc i\kern-.025em b}\kern-.08em
    T\kern-.1667em\lower.7ex\hbox{E}\kern-.125emX}}
\usepackage{balance}
\usepackage{diagbox}
\usepackage[ruled,vlined]{algorithm2e}
\usepackage{tikz}
\usepackage{enumitem}
\usepackage{balance}
\usepackage{booktabs} % For formal tables
\usepackage[caption=false]{subfig}
\usepackage{graphicx}
\usepackage{subfig}
\usepackage{url}
\usepackage{xcolor}
\usepackage{tikz}
\usepackage{adjustbox}
\usepackage{booktabs}
\usepackage{rotating}
\usepackage{multirow}
\usepackage{lineno}
\usepackage{color, colortbl}
\definecolor{grey}{gray}{0.9}
\usepackage{hyperref}
\hypersetup{colorlinks,linkcolor={blue},citecolor={blue},urlcolor={blue}}  
\usepackage[misc,geometry]{ifsym}
\usepackage{dirtytalk}
\usepackage[normalem]{ulem}
\usepackage{blindtext}
\usepackage{tabularx}
\usepackage{tabulary}
\usepackage{wrapfig}
\usepackage{amssymb}
\definecolor{orcidlogocol}{HTML}{A6CE39}
\usepackage{multirow}
\usepackage{float}
\usepackage{svg}
\usepackage{cite}
\newcommand\Tstrut{\rule{0pt}{2.6ex}}       % "top" strut
\newcommand\Bstrut{\rule[-1.5ex]{0pt}{0pt}} % "bottom" strut
\newcommand{\TBstrut}{\Tstrut\Bstrut} % top&bottom struts
\usepackage{physics}
\usepackage[cmintegrals]{newtxmath}
\usepackage{background}
\backgroundsetup{contents=Draft, opacity=0.15, scale=37, color=gray, angle=60}
\begin{document}
%-------------------------------------------------------------------------------

\title{Resisting Deep Learning Models Against Adversarial Attack Transferability via Feature Randomization}
%for single author (just remove % characters)

\author{Ehsan~Nowroozi,~\IEEEmembership{Senior Member,~IEEE,}
        Mohammadreza~Mohammadi,~\IEEEmembership{Member,~IEEE},\\~Pargol~Golmohammadi,~\IEEEmembership{Member,~IEEE},~Yassine~Mekdad,~\IEEEmembership{Member,~IEEE,}~\\Mauro~Conti,~\IEEEmembership{Fellow,~IEEE,}~and~A.~Selcuk~Uluagac,~\IEEEmembership{Senior Member,~IEEE}% <-this % stops a space
\IEEEcompsocitemizethanks{\IEEEcompsocthanksitem E. Nowroozi is with Bahçeşehir University, Faculty of Engineering and Natural Sciences, Computer Engineering Department, Istanbul, Turkey (Email: ehsan.nowroozi65@gmail.com, ehsan.nowroozi@eng.bau.edu.tr)%E. Nowroozi is with the Faculty of Engineering and Natural Sciences (FENS), Center of Excellence in Data Analytics (VERİM), Sabanci University, Istanbul Turkey 34956. E-mail: ehsan.nowroozi65@gmail.com, ehsan.nowroozi@sabanciuniv.edu.
\IEEEcompsocthanksitem M. R. Mohammadi, P. GolMohammadi, and M. Conti are with the Department of Mathematics, Security \& Privacy Research Group (SPRITZ), University of Padua, 35121, Padua, Italy.Email: \{mohammadreza.mohammadi, pargol.golmohammadi\}@studenti.unipd.it, conti@math.unipd.it
\IEEEcompsocthanksitem Y. Mekdad, and A. S. Uluagac are with the Cyber-Physical Systems Security Lab, Department of Electrical and Computer Engineering, Florida International University, Miami, FL, 33174. E-mail: \{ymekdad, suluagac\}@fiu.edu
}% <-this % stops an unwanted space

}

 % end author

\maketitle

\begin{abstract}

In the past decades, the rise of artificial intelligence has given us the capabilities to solve the most challenging problems in our day-to-day lives, such as cancer prediction and autonomous navigation. However, these applications might not be reliable if not secured against adversarial attacks. In addition, recent works demonstrated that some adversarial examples are transferable across different models. Therefore, it is crucial to avoid such transferability via robust models that resist adversarial manipulations.

In this paper, we propose a feature randomization-based approach that resists eight adversarial attacks targeting deep learning models in the testing phase. Our novel approach consists of changing the training strategy in the target network classifier and selecting random feature samples. We consider the attacker with a Limited-Knowledge and Semi-Knowledge conditions to undertake the most prevalent types of adversarial attacks. We evaluate the robustness of our approach using the well-known UNSW-NB15 datasets that include realistic and synthetic attacks. Afterward, we demonstrate that our strategy outperforms the existing state-of-the-art approach, such as the Most Powerful Attack, which consists of fine-tuning the network model against specific adversarial attacks. Finally, our experimental results show that our methodology can secure the target network and resists adversarial attack transferability by over 60\%.

\end{abstract}

\section{Introduction}
The recent studies on the vulnerabilities of machine and deep learning have attracted researcher's attention~\cite{Papernot2016TheSettings,ozposter}. This new field is known as \textit{Adversarial Machine Learning} and has been widely investigated~\cite{Akhtar2018ThreatSurvey,Wang2020AdversarialSurvey,newaz2020adversarial}. In particular, artificial neural networks that are frequently used in deep learning such as Convolutional Neural Networks (CNNs), including image classification, computer vision, network security~\cite{naseem2021minos}, and natural language processing, has raised significant concerns about the vulnerability of these models to adversarial attacks. To that end, we distinguish two different attack strategies in machine learning techniques: poisoning attacks and exploratory evasion attacks. In the poisoning attack, the adversary is aware of the training examples, and the attack is performed throughout the training process, while in the exploratory evasion attack, the adversary compromises the network during the testing phase.

To the best of our knowledge, most machine and deep learning techniques are intrinsically vulnerable to various types of adversarial attacks~\cite{Nowroozi2021AForensics,nowroozi2022adversarial_URL}. Moreover, the adversary can transfer some of these attacks from one network (a.k.a, Source Network (SN)) to another one (a.k.a, Target Network (TN)). As a result, it is necessary to consider preventing this behavior since the transferability property between the networks presents a challenging security issue~\cite{Wang2019AssessingClassifiers,Barni2019OnForensics}. Therefore, it is crucial to consider from a research point of view preventing such transferability, and thus by improving the TN's security against adversarial attacks.

In this paper, we are motivated to provide a new model to secure the TN and avoid adversarial transferability between the SN and the TN. More specifically, we demonstrate that our model can significantly improve TN's security. Our main contribution consists of strengthening the TN against exploratory evasion attacks. In particular, we modify the TN's classifier by considering a Feature Randomization (FR) strategy to extract random features from the SN's flatten layer. In our work, we experimentally demonstrate the transferability of eight adversarial attacks (with different parameters) from the SN to the TN. Then, we improve the TN's security via an FR technique by considering two adversarial settings: the case where the adversary has a Limited Knowledge (LK) about the TN, and the case where the adversary has a Semi Knowledge (SK) about the TN. Moreover, we show the performance of our strategy against the MPA approach. For the replication of our experiments, we made our implementation code publicly available~\cite{EhsanGitHub}.

\noindent \textbf{Summary of Contributions.} The major novelty and contributions of our work are summarized as follows:
\begin{itemize}
    \item We propose a novel FR approach that improves the target network's security against eight adversarial attacks in both LK and SK conditions. We focus on evasion attacks in the testing phase that try to decrease the false positive rate, i.e., detecting attacked samples as pristine.
    \item  We demonstrate that adversarial attacks may be transferred across the SN and the TN, emphasizing the importance of developing robust models against this property. Our results show that for both shallow/deep networks, most of the considered attacks have an Attack Success Rate (ASR) of more than 95\%. 
    \item We design and construct a Support Vector Machine (SVM) in the TN based on the FR strategy and test its robustness under different adversarial settings. By analyzing a wide range of random feature vectors, we find that the feature vectors of sizes 30,50, and 200 achieve a promising secure model. 
    \item We evaluate the effectiveness of the FR strategy against different adversarial attacks, including the I-FGSM, the FGSM, the BIM, the PGD, the L-BFGS, the JSMA, the DeepFool, and the C\&W attack. The experimental results of our study shows that the FR methodology is more efficient than the MPA approach, and can defeat adversarial transferability over 60\%.
    
\end{itemize}

\noindent \textbf{Organization.} In Section~\ref{BKG}, we overview the foundation of adversarial attacks against machine and deep learning models. Then, we describe the considered datasets. We discuss the related works in Section~\ref{RW}. Section~\ref{Problem} presents the problem scope and threat model. In Section~\ref{Methodology}, we describe our proposed approach to improve the security of the TN. In Section~\ref{Attacks}, we report our adversarial attacks with their corresponding attack parameters. We propose the defense mechanisms in Section~\ref{Defenses}, and Section~\ref{Discussion} discusses the outcomes of our approach. Finally, we conclude the paper and present future work in Section~\ref{Conclusion}.

\section{Background}
\label{BKG}
In this section, we provide the background on adversarial attacks on deep learning-based models. Table ~\ref{Abbreviations} provides a list of the acronyms and abbreviations used in this study.

\begin{table}[!h]
%\scriptsize
%\footnotesize
\centering
\small
\caption{Acronyms and Abbreviations List. \label{Abbreviations}}
\begin{tabular}{@{}|c|l|}
\hline
\textbf{Acronym} & \textbf{Description} \\ \hline

CNN & Convolutional Neural Network \\ \hline

FR & Feature Randomization \\ \hline

LK & Limited Knowledge \\ \hline

SK & Semi Limited Knowledge \\ \hline

PK & Perfect Knowledge \\ \hline

SVM & Support Vector Machine \\ \hline

ML & Machine Learning \\ \hline

DL & Deep Learning \\ \hline

SN & Source Network\\ \hline

TN & Target Network\\ \hline

$N_1$, $N_2$ & Shallow Network, Deep Network \\ \hline

ASR & Attack Success Rate \\ \hline

Max. dist & Maximum distortion\\ \hline

PSNR & Peak signal-to-noise ratio\\ \hline

$L_1$ dist & $L_1$ distance\\ \hline

I-FGSM & Iterative-Fast Gradient Sign Method \\ \hline

PGD  & Projected Gradient Descent\\ \hline

JSMA  & Jacobian Saliency Map Attack\\ \hline

BIM & Basic Iterative Method \\ \hline

C\&W & Carlini \& Wagner \\ \hline

L-BFGS  &  Limited-memory Broyden-Fletcher-Goldfarb-Shanno\\ \hline

FGSM & Fast Gradient Sign Method\\ \hline

MPA & Most Powerful Attacks \\ \hline

Lr & learning rate \\ \hline
\end{tabular}
\end{table}

According to the Deep Learning literature, the attacker can fool CNN-based models in three different settings: white-box, gray-box, and black-box attacks. For each of these settings, the adversary has information about the TN with different levels of knowledge. These settings are classified into three categories: Limited Knowledge (LK), Semi Knowledge (SK), and Perfect Knowledge (PK).
\begin{itemize}
    \item \textit{\textbf{LK:}} Commonly known as black-box attacks, and indeed regarded where the attacker cannot access the hyper model parameters, which is a more complex scenario in digital forensics; as a result, the attacker conducts multiple queries to obtain the internal details of the model.
    \item \textit{\textbf{SK:}} In this scenario, the adversary has a partial information about the victim's network and performs his attacks under gray-box settings ~\cite{NEURIPS2018}.
    \item \textit{\textbf{PK:}} It occurs when an adversary has full knowledge of the forensic algorithm; hence, this scenario is referred to as a white-box setting and considered the ideal scenario for the adversary~\cite{Zhang2020AdversarialChallenges}.
\end{itemize}

In our study, we launched eight adversarial attacks with different parameters in black-box settings without accessibility to the model, namely: The JSMA \cite{Papernot2016TheSettings}, the PGD  \cite{Madry2018}, the L-BFGS  \cite{Szegedy2013IntriguingNetworks}, the I-FGSM  \cite{Kurakin2017ADVERSARIALWORLD}, the FGSM ~\cite{GoodfellowEXPLAININGEXAMPLES}, the DeepFool ~\cite{Moosavi-DezfooliDeepFool:Networks}, the BIM ~\cite{Kurakin2017ADVERSARIALWORLD}, and the C\&W attacks~\cite{carlini2017towards}. %(with the strength parameters 0 and 100). 

\subsubsection{\textbf{The I-FGSM attack}}%%%%%%%%%%%%%%%%%%%%%%%%%%%%%%%%%%%%%%%%%%%%%%%%%%%%%%%%%%%%%%%%%%%%%%%%%%%%%%%%%%%%%%%%%%%%%%%%%%%%%%%%%%%%%%%%%%%%%%%%%%%%%%%%%%%%%%%%%%%%%%%%%%%%%%%%
This attack can be described as the iterative method of the FGSM attack, which is a gradient method that relies on maximizing the loss function by adjusting the input data. Subject to an upper bound on the perturbation, the I-FGSM attack aims to make the classifier perform inadequately~\cite{GoodfellowEXPLAININGEXAMPLES}. The I-FGSM attack is one of the most popular adversarial attacks designed to target deep neural networks, and which consequently leads to misclassify the data input. The FGSM attack highlights major issues with the potential unboundedness of the perturbed data $S$~\cite{GoodfellowEXPLAININGEXAMPLES}. This issue can be addressed by the I-FGSM method. It constitutes a bound constraint on $S$, which uses an iterative linearization rather than the
one-shot linearization in the FGSM attack~\cite{GoodfellowEXPLAININGEXAMPLES}. For $S$, an input of the model with a ground label truth $Z$ and $\theta$ parameters, the adversarial sample $Adv_{S}$ of the FGSM attack is expressed as follows:  
\begin{equation}
Adv_{S} = \alpha \times sign(\triangledown_{S}I(\theta,S,Z)) + S.
\end{equation}
$I$ is the cross-entropy function, and $\alpha$ is the factor responsible for normalizing the attack strength. $\alpha$ is the variable that controls the perturbations and must be minimal to enable the adversarial attack. The iterative variant of the FGSM approach \cite{Kurakin2017ADVERSARIALWORLD} has better perturbations than the FGSM strategy and may be considered as its extension. In particular, we iterate the FGSM method to the gradient sign with smaller variations. In the I-FGSM attack, and for $Adv^{0}_{S} = S$, we compute the adversarial sample for each variation $i+1$ as follows: 
\begin{equation}
        Adv^{i+1}_{S} = \alpha * sign(\triangledown_{S}I(\theta,Adv^{i}_{S},Z)) + Adv^{i}_{S}.
\end{equation}
We refer to the I-FGSM attack with the parameter $\alpha = 10$ as I-FGSM010.

\subsubsection{\textbf{The JSMA attack}}%%%%%%%%%%%%%%%%%%%%%%%%%%%%%%%%%%%%%%%%%%%%%%%%%%%%%%%%%%%%%%%%%%%%%%%%%%%%%%%%%%%%%%%%%%%%%%%%%%%%%%%%%%%%%%%%%%%%%%%%%%%%%%%%%%%%%%%%%%%%%
Given the high number of perturbations caused by the I-FGSM attack, the authors in~\cite{Papernot2016TheSettings} proposed the JSMA attack to reduce such perturbations. As a result, the JSMA-based adversarial samples exhibit fewer perturbations than the I-FGSM attack and are significantly more difficult to detect for targeted misclassification. The JSMA attack is a gradient-based approach that utilizes adversarial saliency maps and a forward derivative method. For a model with $N$-dimensional input $S$, the adversary approximates the Jacobian matrix of the classifier $F$ learned during the training phase. The matrix has $M$-dimension and is expressed by:
\begin{equation}
        J_{F}(S) = \pdv{F(S)}{S} = {\begin{bmatrix} 
        \pdv{F_{j}(S)}{s_i}.
        \end{bmatrix}}_{i \in 1..M, j \in 1..N}.
\end{equation}
Afterward, the adversary constructs the adversarial saliency maps using the forward derivative method. These maps determine the features that enable the perturbations for the intended output. Therefore, the adversary builds the space of adversarial samples. The adversarial saliency map $U$ for a target class $o$ is defined as follows:
\begin{equation}
U(S, o)[i]=\left\{\begin{array}{ll} 0, if \frac{\partial F_{o}}{\partial s_{i}}(S)<0 \text { or } \sum_{j \neq o} \frac{\partial F_{j}}{\partial s_{i}}(S)>0. & \\ {\frac{\partial F_{o}}{\partial s_{i}}(S)\left|\sum_{j \neq o} \frac{\partial F_{j}}{\partial s_{i}}(S)\right|,}otherwise. & \end{array}\right.
\end{equation}
To summarize, JSMA approach relies on forwarding propagation to produce a saliency map for each iteration and indicates the data that emphasize the categorization. At a high value, for example, implies that altering the data can increase the possibility of misclassification. According to the map, the data are individually changed by a parameter {$\theta$} (i.e., it defines the range where the data have changed). We refer to the JSMA attack with the parameter $\theta = 1$ as JSMA001. 

\subsubsection{\textbf{The L-BFGS attack}}%%%%%%%%%%%%%%%%%%%%%%%%%%%%%%%%%%%%%%%%%%%%%%%%%%%%%%%%%%%%%%%%%%%%%%%%%%%%%%%%%%%%%%%%%%%%%%%%%%%%%%%%%%%%%%%%%%%%%%%%%%%%%%%%%%%%%%%%%%%%%%%%%%%%%%%%%%%%
It is a nonlinear gradient strategy that involves increasing the prediction error and optimizing the input data in order to generate adversarial examples~\cite{Szegedy2013IntriguingNetworks}. 
The L-BFGS approach can be considered as a box-constrained optimization methodology for generating adversarial examples, with the same intention of minimizing perturbations for a sample input. Given a classifier $F$, a minimizer $r$, a target label $l$, and $S$ as inputs, the L-BFGS attack is formally expressed as:
\begin{equation}
\label{Eq}
\min_{Adv_{S}} || S - Adv_{S} ||^{2}_{2}  \mbox{  subject to: } F(S+r) = l.
\end{equation}
However, the problem presented in Eq.~\ref{Eq} is highly non-linear and time consuming. The following equation is an approximate approach to solve the optimization problem based on the usage of box-constrained L-BFGS: 
\begin{equation}
\min_{Adv_{S}} c.|| S - Adv_{S} ||^{2}_{2} - I(\theta,S,Z). 
\end{equation}
The above-mentioned estimation is used to determine the minimum positive scalar $c$ that satisfies the minimizer $r$ in the Eq.~\ref{Eq}.

\subsubsection{\textbf{The PGD attack}}%%%%%%%%%%%%%%%%%%%%%%%%%%%%%%%%%%%%%%%%%%%%%%%%%%%%%%%%%%%%%%%%%%%%%%%%%%%%%%%%%%%%%%%%%%%%%%%%%%%%%%%%%%%%%%%%%%%%%%%%%%%%%%%%%%%%%%%%%%%%%%%%%%%%%%%%%%%%%%
This attack provides adversarial examples by employing local first-order over the TN \cite{Madry2018}. It is an iterative and first-order attack that employs uniform random noise for the initialization. Differently from the I-FGSM attack, the input $X$ is updated for each variation $i+1$ according to the following rule:
\begin{equation}
 Adv^{i+1}_{S} = \Omega_{S+Q}(\alpha * sign(\triangledown_{S}I(\theta,S,Z)) + Adv^{i}_{S}).
\end{equation}
The projection operator $\Omega$ holds $Adv^{i+1}_{S}$ within a range of perturbations $Q$. The PGD approach considers the $L_\infty$ distortion and looks for perturbations to strengthen $I(\theta,S,Z)$. We refer to the PGD attack with the parameter $\alpha = 5$ as PGD005. 

\subsubsection{\textbf{The DeepFool attack}}%%%%%%%%%%%%%%%%%%%%%%%%%%%%%%%%%%%%%%%%%%%%%%%%%%%%%%%%%%%%%%%%%%%%%%%%%%%%%%%%%%%%%%%%%%%%%%%%%%%%%%%%%%%%%%%%%%%%%%%%%%%%%%%%%%%%%%%%%%%%%%%%%%%%%%%
This adversarial attack aims to fool multiclass classifiers with minimal possible perturbations~\cite{Moosavi-DezfooliDeepFool:Networks}. More specifically, the DeepFool attack approximates the decision space of the classifier in order to find the minimal perturbations. For a classifier, we formally express the minimal perturbation needed to produce an adversarial sample by:
\begin{equation}   
 \delta(S,F) = \min_{r}||r||_{2} \mbox{ subject to: } F(S+r) \neq F(S). 
\end{equation}
We refer to the robustness of $F$ for the input $S$ by $\delta$ and the minimal perturbation by $r$. The generated data by DeepFool contains the lowest possible amount of noise which is sufficient to fool the neural network model into classifying it as a sample of another class \cite{deepfoolcite}.

\subsubsection{\textbf{The Carlini \& Wagner attack}}
The Carlini \& Wagner (C\&W) attack generates high-confidence adversarial examples by accessing the parameters and architecture of the network~\cite{carlini2017towards}. However, it has a high-cost generation of adversarial examples. The C\&W attack methodology can be performed under three different attack scenarios that can be referred to as distance metrics: $L_{2}$, $L_{0}$, and $L_{\infty}$. In the $L_{2}$ attack, given a benign sample $s$ and a chosen target class $t$ different from the benign class of $s$ ($t \neq C^{*}(x) $), the attacker's goal is to search for the value $w$ minimizing the following expression:

\begin{equation}
 \min_{w}||\frac{1}{2}(\tanh{(w)}+1)-x||^{2}_{2} + c . f(\frac{1}{2}(\tanh{(w)}+1)) . 
\end{equation}

Where $c$ is an acceptable constant and $f$ is a function defined as follows:

\begin{equation}
f(x')=max(max\{P(x')_{i} : i \neq t\} - P(x')_{t}, -T).
\end{equation}

We state $P$ as the outcome of all layers, excluding the softmax layer, and $T$ as a parameter that may be used to control the degree of adversarial examples. The $L_{0}$ attack is the iterative version of the $L_{2}$ attack. It consists of computing the gradient of $f$, and evaluates the adversarial sample of the $L_{2}$ attack. In other words, for a solution $\delta$ returned from $L_{2}$ attack regarding a benign sample $s$, we compute its gradient $g$ that is defined as follows:
\begin{equation}
g = \triangledown f(s+\delta) .
\end{equation}
It is worth mentioning that the $L_0$ is more challenging than the $L_2$ attack. The $L_\infty$ attack relies on naively minimizing the following equation to generate adversarial examples: 
\begin{equation}
 \min_{\delta} c . f(s+\delta) + ||\delta||_{\infty}. 
\end{equation}
In our work, we refer to the C\&W attack with the strength parameter $0$ and $100$ by CW0 and CW100, respectively.

\section{Related work}
\label{RW}
Several studies in the literature proposed adversarial attacks and countermeasures for deep learning-based networks \cite{OH2022511, cite-key,Anelli2022}. In this section, we overview the prior works on countermeasures techniques in deep learning models that leverage the randomization strategy. Then, we outline the differences from existing works.

Recent works proposed adding randomization layers to improve the robustness of CNN-based models. In~\cite{ashrafiamiri2020r2ad}, the authors presented a two-layer defense-based neural network; the first layer is a random nullification layer that consists of randomly deleting some features from the input to minimize the adversarial perturbations, while the second layer is an auto-encoder-based reconstructor that rebuilds the input features and performs the classification tasks. The numerical results against the FGSM, BIM, JSMA, DeepFool, and C\&W attacks show high robustness with an accuracy of up to 80\%. Similarly, the authors in~\cite{xie2017mitigating} proposed randomization at inference time technique to defeat iterative adversarial attacks. This technique adds two randomization layers at the beginning of the classification neural network. The first layer consists of a random resizing layer, which resizes the input samples randomly. The second layer performs random padding operations for the samples with zeros before its transformation to the CNN model. The experiments performed with such a randomization strategy demonstrate its robustness against the FGSM, the DeepFool, and the C\&W attacks. However, additional computations are required by adding the random resizing and random padding layers. 

Another approach is developed by Taran et al. ~\cite{taran2019defending}, the authors employ a randomized diversifying mechanism to protect neural networks against various attacks in classification. Such a strategy is implemented in a multi-channel architecture and utilizes a shared secret key between the training and testing stages. The experimental results regarding the randomization diversification mechanism show the robustness against the C\&W attack with different parameters.

The authors in ~\cite{eniser2020raid} presented the Randomized Adversarial-Image Input Detection (RAID) for Neural Networks, which relies on building a secondary classifier capable of detecting malicious input based on neuron activation values. The number of monitoring neurons is randomly selected and labeled as activation fingerprints. The authors evaluated the effectiveness of RAID against six adversarial attacks: the PGD, the FGSM, the BIM, the DeepFool, the C\&W, and the JSMA attack. The experimental results demonstrate a good detection accuracy of up to 90\% for different attacks.

\noindent
\textbf{Differences from existing works.} Different from existing studies, our approach improves the security of the TN and avoid the transferability property by considering eight adversarial attacks. Moreover, our feature randomization strategy relies on changing the TN's classifier into a Support Vector Machine. Consequently, we decrease the attacker's knowledge by changing the architecture of the target's network. In table~\ref{difference}, we show the difference between our work and existing works. 

%Several studies in the literature proposed adversarial attacks and defenses for deep learning-based models \cite{OH2022511, cite-key,Anelli2022}. In this section, we overview prior works on defense approaches in deep learning models that leverage the randomization strategy.
%In another work proposed by Taran et al.~\cite{taran2019defending}, the authors utilize a randomized diversification strategy to protect the neural networks against adversarial attacks in classification systems. 
%In~\cite{eniser2020raid}, the authors introduced RAID: Randomized Adversarial-Image input Detection approach for neural networks. This approach relies on training a secondary classifier to identify malicious input through neuron activation values. 
%
\begin{table*}[!h]
\centering
\small
\caption{Difference between our work and existing works} \label{difference}
%\begin{tabular}{|l|l|l|l|l|l|l|}
\resizebox{1\textwidth}{!}{
\begin{tabular}{|l |l |l |l |l |l | l |}
\hline
\centering
\textbf{Ref.} & \textbf{App Domain} & \textbf{Attacks} & \textbf{Datasets} & \textbf{Pros} & \textbf{Cons}    \\ \hline

    \cite{ashrafiamiri2020r2ad} & \makecell[l]{Computer vision} & \makecell[l]{FGSM, JSMA,  \\ BIM, Deepfool, \\and CW. } &  \makecell[l]{-MNIST \\ -Fashion-MNIST} & \makecell[l]{-Minimize the effect of adversaries \\ for the trained network. \\ 
    -High robustness against adversaries\\
     and higher performance on normal \\ samples.}& \makecell[l]{-Low performance in CW which is close \\to 70\%. \\ -Low performance in black-box attacks \\ on Fashion-MNIST dataset e.g. FGSM \\ and BIM.}          \\ \hline
    
    \cite{xie2017mitigating} & \makecell[l]{Computer vision} & \makecell[l]{FGSM, Deepfool, \\and CW.} &  \makecell[l]{ImageNet} &\makecell[l]{-adversarial examples rarely transfer \\ for iterative attacks. \\- High accuracy on clean examples.} & \makecell[l]{-Low performace for the networks \\ Inception-v3 and ResNet-V2, \\ i.e. random brightness, and brightness++.}                       \\ \hline
    
    \cite{taran2019defending} & \makecell[l]{Computer vision} & \makecell[l]{CW} &  \makecell[l]{-MNIST\\-Fashion-MNIST\\-CIFAR-10} &  \makecell[l]{-Reduce back gradient propagation.}& \makecell[l]{-Failed gradient sparse, and non-gradient \\based attacks.}                         \\ \hline
    
    \cite{eniser2020raid} & \makecell[l]{Computer vision} & \makecell[l]{
PGD, FGSM, \\BIM, DeepFool,\\CW, and JSMA} &  \makecell[l]{-MNIST \\-CIFAR-10} & \makecell[l]{-90\% accuracy with the strongest \\attacks (CW, DF), and excellent\\ detection versus weaker adversaries\\ (i.e., PGD, BIM, and FGSM).}  & \makecell[l]{-Need to run the tool through its tests \\with additional threat models, wider neural \\networks, and diverse\\ tasks like natural language processing.}                         \\ \hline
    
 Our work. & \makecell[l]{Computer networks.} & \makecell[l]{
I-FGSM, FGSM, \\BIM, PGD, L-BFGS,\\ JSMA, DeepFool, \\ and CW } &  \makecell[l]{-UNSW-NB15} & \makecell[l]{-Improve a security in a computer\\ networks domain in Lk and SK. \\-Avoid a transferability issue\cite{EhsanDemystifying}.\\-considering different ML and DL\\ models.}  & \makecell[l]{-\textbf{Need to investigate:} \\ Attack transferabilily in poisioning \\ attacks, and backdoor attacks.}                       \\ \hline
  
\end{tabular}}
\end{table*}

\section{Problem Scope and Threat Model}
\label{Problem}
The transferability property is satisfied if the adversarial samples that compromises the SN can be used to fool the TN. In this case, we provide the assumptions regarding the attacker's capabilities. In what follows, we outline the scope of the problem addressed in this paper and the threat model. 

\subsection{Problem Scope}

\begin{figure}[!h]
\begin{center}
\includegraphics[width=0.33\textwidth]{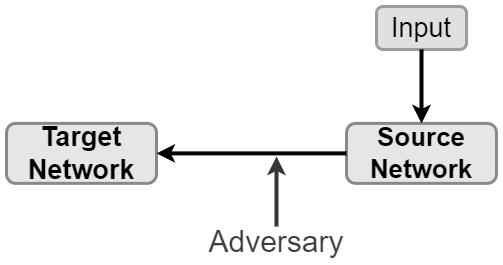}
\caption{Visual representation of the transferability property in Convolutional Neural Networks.}
\label{trans}
\end{center}
\end{figure}

In computer networks, resisting adversarial sample attacks on a network traffic is challenging. Hence, extensive research efforts have been conducted to design secure and robust deep neural networks. In particular, prior works studied the capabilities of transferring adversarial samples from the SN to the TN. This property is known as \textit{transferability}, and is depicted in Fig.~\ref{trans}. In this work, we focus on network traffic classification, adversarial sample attacks, and defense mechanisms.

In \cite{Dmystify}, the authors demonstrated that most adversarial attacks could not be transferable between SN and TN, but only a few hold the transferability property. Therefore, in security-oriented applications that consider machine/deep learning models, it is crucial to implement strong defense mechanisms to avoid such transferability between the SN and the TN. In this line of research, TN's security can be improved using two approaches. The first approach consists of resisting the TN by fine-tuning it with the MPA. In contrast, the second approach relies on decreasing the attacker's knowledge of the TN (i.e., putting the attacker in the LK condition) through an FR strategy. In the first approach, we fine-tune the TN with eight adversarial attacks to get eight different tuned models. Then, we test the fine-tuned models against the I-FGSM, the FGSM, the JSMA, the BIM, the PGD, the L-BFGS, the DeepFool, and the C\&W attacks. In the second approach, we select random features from the flattened layer of the CNN network to decrease the attacker's knowledge and feed them to the SVM to classify them.

\subsection{Threat Model}

In real-world applications, the attacker has a very limited knowledge to the TN (refer to as black-box setting). For this reason, we consider our experiments under black-box settings. In this case, with the limited capabilities of the adversary that cannot have access to the TN (e.g., parameters of the model, network architecture), we assume that the adversary has a LK and can potentially increase his access to the victim's network to get into an SK scenario. More specifically, the attacker builds the adversarial examples on the SN, which is trained using a different or similar dataset from the TN. Then, the adversary launch different adversarial attacks to fool the TN. In this study, we test and develop a generalized approach for black-box attacks against DL models that take advantage of adversarial example transferability. 

%In this section, we introduce the problem scope and the threat model. In particular, The \textit{transferability} is defined as whether the adversarial samples generated based on SN can be used to attack TN. If the attack is successful, we can confirm that the attacks are transferred from the source to the target network. Then, we provide the assumptions regarding the attacker's capabilities.
%In~\cite{Dmystify}, the authors demonstrated the non-transferability of most attacks between the SN and the TN. However, some adversarial attacks with specific parameters enable a transferability property. 
%In the Deep Learning literature, the attacker can fool the CNN-based models under three settings: black-box, gray-box, and white-box attacks.

%Since the attacker has complete access to the TN in the white-box scenario; it is important to consider the black-box scenario given its practicality in real-world applications. 
\section{Methodology}
\label{Methodology}
In this section, we introduce our novel method to improve TN's security by including a feature randomization strategy to mitigate dangerous adversarial attacks. In what follows, we describe the considered datsets, the learning models, their parameters, as well as the shallow ($N 1$) and deep network ($N 2$) architectures.

\subsection{Proposed Approach}

\begin{figure*}
    \centering
    \includegraphics[width=0.65\textwidth]{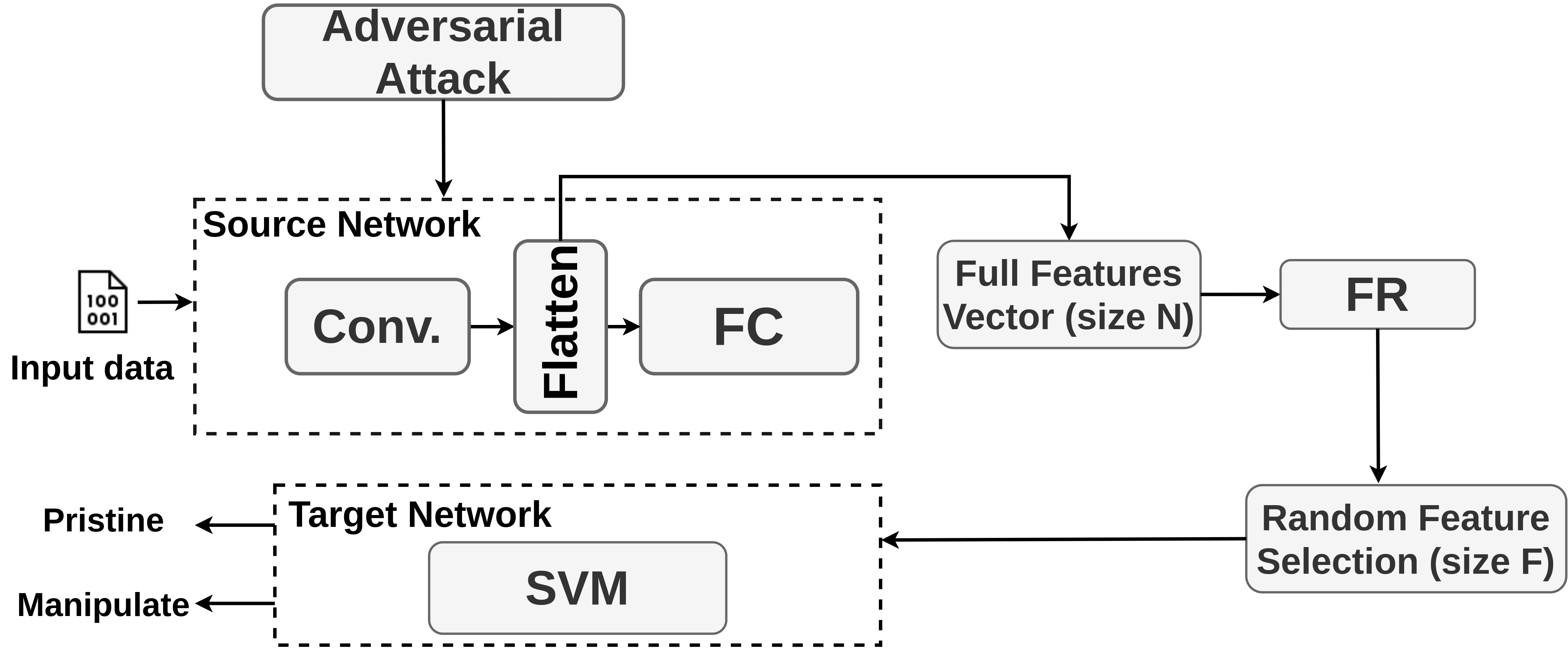}
    \caption{\centering Proposed approach to improve the TN's security}
    \label{fig:FR_method}
\end{figure*}

To decrease the adversary's knowledge, we propose an FR strategy by changing the TN's classifier into a Support Vector Machine (SVM), as depicted in Figure.~\ref{fig:FR_method}. For the SN, we consider a Convolutional Neural Network (CNN), while for the target network, we consider the SVM classifier that receives random features from the flattened layer of the CNN. The flatten layer is a layer that is usually placed after convolutional layers and before dense layers or classification layers. The flattened layer receives multi-dimensional input and gives a single-dimensional vector as the output, which is the suitable data format to consider in classifiers like SVM and fully-connected networks. In the FR strategy, we flatten the output of the final layer of CNN in order to get a one-dimensional output. Then, we utilize this output as input for the randomization procedure, and we specifies a random amount of it that will be given as input to the SVM for classification.

\subsection{Network Architecture and Learning Models}

Due to the wide usability of the networks proposed by Bayer et al.~\cite{bayar2018constrained} and Barni et al.~\cite{Barni2018Cnn-basedPost-processing} for many applications such as multimedia forensics, computer vision, cybersecurity, and, in particular, network security \cite{Barni2019OnForensics} \cite{Demystifying}, we believe that these networks are suitable for our investigations and consider them for constructing the shallow($N_1$) and deep($N_2$) network. To demonstrate the applicability of our approach on both $N_1$ and $N_2$, all the processes have been done separately, and these two networks are not connected. In what follows, we describe the $N_1$ and $N_2$ respectively. It is worth mentioning that $N_1$ and $N_2$ are SN and should not be considered as a TN. 

\subsubsection{\textbf{Shallow Network ($N_1$)}}

For the shallow network $N_1$, we consider the architecture presented in~\cite{bayar2018constrained}. It consists of 3 convolutional layers, namely constrained convolutional layers. This architecture can adaptively learn manipulation detection features directly from the data with high accuracy. Moreover, it outperforms existing image manipulation detection techniques~\cite{qiu2014universal}, especially when considering real large-scale training datasets. Therefore, we can perfectly use this model as a forensic detector for different image manipulation.

\subsubsection{\textbf{Deep Network ($N_2$)}}

For the $N_2$, we consider the architecture proposed in~\cite{Barni2018Cnn-basedPost-processing}. This network relies on eight convolutional layers and can be seen as a patch-based CNN. This network can detect contrast-adjusted images with a good performance in the presence of JPEG post-processing operations. Additionally, it achieves high accuracy under different Quality Factors (QFs).

\subsubsection{\textbf{Description of the datasets}}

In our study, we consider the UNSW-NB 15 \cite{unswdataset} dataset generated by the IXIA PerfectStorm application to generate a combination of realistic modern routine operations and synthetic existing attack characteristics. The raw network packets of this dataset are obtained via the tcpdump tool to capture 100 GB of the raw traffic (e.g., Pcap files). The eight kinds of attacks in this dataset include Backdoor attacks, traffic analysis, Fuzzers, DoS attacks, Generic traffic, Exploits, Reconnaissance, Worms, and Shellcode. The total of items is 2540044, which is divided into four CSV files and merged into one file.

In the preprocessing phase of the datasets, we initially have 48 features. Some of these features are not effective for analyzing the traffic in the process of the attacks (e.g., backdoor attacks, worm attacks). These features are the source port, source IP, destination port, destination IP, state, and protocol. After removing these features, we got 42 features. Then, we constructed matrices where the product of their rows and columns is equal to 42. In this case, the matrices have the size of $6 \times 7$. Given the small size of these matrices that cannot be used for CNNs to distinguish between pristine and manipulated samples, we reshaped the size of these matrices from $6 \times 7$ to $64 \times 64$. In our final dataset, we stored around 319.480 pristine images and 319480 manipulated images. Finally, we divide these samples into three categories: the training, validation, and testing samples. 

\subsubsection{\textbf{Description of the Learning Models}}

For the learning models, i.e., $N_{1}$ and $N_{2}$, we used in our experiments 447.266 training samples (223.633 for manipulation and 223633 for pristine), 127.786 validation samples (63.893 for manipulation and 63.893 for pristine), and 63.890 testing samples (31.945 for manipulation and 31.945 for pristine). In Table~\ref{learning}, we provide a numerical description of the learning models on the SN. The training epochs are set to 20 for $N_1$ and 10 for $N_2$, and we considered Adam optimizer with a learning rate (Lr) of $10^{-6}$ for $N_1$ and $10^{-4}$ for $N_2$. We achieved a testing and validation accuracy of more than 95\% for both networks.

\begin{table}[!h]
\centering
\scriptsize
\caption{Description of the Learning Models on the SN \& TN}\label{learning}
\begin{tabular}{|c|c|c|}
\hline
---- & $N_{1}$ & $N_{2}$ \\ \hline
 \# of Conv. Layers & 3 & 9  \\ \hline
 \# of Epochs & 20 & 10 \\ \hline
 \# of Train Batch& 64 & 16 \\ \hline
 \# of Validation Batch& 100 & 16 \\ \hline
 \# of Test Batch& 100 & 100 \\ \hline
 Optimizer& Adam, Lr=1e-06 & Adam, Lr=1e-04 \\ \hline
 Validation Accuracy& 95.99\% & 96.42\% \\ \hline
 Test Accuracy& 95.86\% & 96.42\% \\ \hline
\end{tabular}
\end{table}

%In this section, we present our proposed approach to improve TN's security by considering FR to mitigate dangerous adversarial attacks. 
%Then, we describe the network architecture (i.e., the shallow network ($N_1$) and the deep network ($N_2$)) as well as the learning models and their parameters.
%
%\begin{figure*}[!h]%
%\centering
%	\subfloat[\small Pipeline of the source networks ($N_{1}$ and $N_{2}$) ]{{\includegraphics[width=0.9\textwidth]{Figures/SN_N1N2.png} }}%
%	\\
%	\subfloat[\small Pipeline of the target network (Support Vector Machine)]{{\includegraphics[width=0.4\textwidth]{Figures/TN_SVM.png} }}%
%	\caption{Representation of the considered SN and TN.}%
%	\label{Networks}%
%\end{figure*}	
%
%Training epochs are set to 20 for $N 1$ and to 10 for $N 2$.%The number of training epochs is set to 20 for $N_{1}$ and 10 for $N_{2}$. 
%For optimizing both $N_1$ and $N_2$, we considered Adam optimizer with a learning rate (Lr) of $10^{-6}$ for $N_{1}$ and $10^{-4}$ for $N_{2}$. 
\section{Adversarial Attacks}
\label{Attacks}
In this section, we report about the experiments regarding the adversarial attacks when $N_{1}$ and $N_{2}$ are considered as SN. Then, we present our experimental results for the $N_{1}$ and $N_{2}$ networks when they are considered as TN.

\subsection{Attack Parameters}

In general, we define two types of adversarial attacks against deep learning CNN models: targeted and untargeted attacks. The untargeted attacks enable the trained network to misclassify the input regardless of the output label. In contrast, the targeted attacks aim to deceive a deep learning model by encouraging the model to produce a specific target label for the adversarial sample. For a binary classification task (i.e, which is the case of our study), and since we consider two classes (pristine/manipulate), the targeted and untargeted attacks are similar~\cite{phdthesisehsan}. Therefore, to apply the attacks on $N_{1}$ and $N_{2}$, we perform eight adversarial attacks with different parameters. Each parameter has a different role for different attacks and can influence the performance of the attacks in terms of fooling the DL models. The considered attacks are the most popular adversarial attacks in deep learning: The I-FGSM, the FGSM, the BIM, the PGD, the L-BFGS, the JSMA, the DeepFool, and the C\&W attack. To produce the attack samples, we selected 500 samples randomly from a malicious test dataset; it is obvious that if we select samples from the training dataset, we cannot fool the models, as the networks and models that have considered these data can detect them easily. Afterward, we applied the abovementioned adversarial attacks on the selected samples using the Toolbox library \cite{FoolBox}. Then, we fed these samples to $N_1$ and $N_2$ in order to measure the success rate of each attack. We mention that the 500 selected samples are different for $N_{1}$ and $N_{2}$, and were randomly selected.

\subsection{Experimental Results}
We consider several parameters during our experiments to evaluate our models. In particular, we compute the PSNR, $L_1$ distortion, maximum absolute distortion, and Attack Success Rate (ASR) averages for each of the eight considered adversarial attacks. We define ASR as $m/n$ where $m$ is the number of attacked samples that successfully fooled the model, and $n$ is the number of all the attacked samples. Then, we report the results in Table~\ref{attack_res_n1} and Table~\ref{attack_res_n2} for the $N_{1}$ and $N_2$ when considered as TN.

\subsubsection{\textbf{Experimental Results on the $N_1$}}

In $N_{1}$ with test accuracy of 95.86\%, we remark that the average PSNR is less than 46dB, and more than 80\% of adversarial attacks succeeded with a high ASR. Given the number of convolutional layers in $N_{1}$, the reported experimental results in Table~\ref{attack_res_n1} are expected. This could be explained due to the inherent vulnerability of machine learning models.

\begin{table}[!h]
\centering
\scriptsize
\caption{Attack Results on the $N_1$ when considered as TN. \label{attack_res_n1}}
\resizebox{1\columnwidth}{!}{
\begin{tabular}{|c|c|c|c|c|}
\hline
\textbf{Attack Type} & \textbf{PSNR} & \textbf{$L_1$ dist} & \textbf{Max. dist} & \cellcolor{lightgray} \textbf{ASR}   \TBstrut\\ \hline

 I-FGSM, $\varepsilon$ = 0.1 & 36.4225 & 3.0011 & 6.0792 & \cellcolor{lightgray} 1.00 \TBstrut\\ \hline

 FGSM, $\varepsilon$ = 0.1 &  9.4111 & 80.1985 & 143.4194 & \cellcolor{lightgray} 0.96 \TBstrut\\ \hline

 JSMA, $\theta$ = 0.01 &  39.9512 & 0.6803 & 17.85 & \cellcolor{lightgray} 0.72 \TBstrut\\ \hline

 BIM, $\varepsilon$ = 0.01 & 18.3842 & 30.0737 & 45.9516 & \cellcolor{lightgray} 0.97 \TBstrut\\ \hline

 LBFGS, $\varepsilon$ = 1e-5 & 46.3229 & 0.7998 & 9.4893 &  \cellcolor{lightgray} 0.99 \TBstrut\\ \hline
 
 DeepFool, default parameter & 45.0145 & 0.897 & 11.8975 & \cellcolor{lightgray} 0.38 \TBstrut\\ \hline

 PGD, $\varepsilon$ = 0.05, step size = 0.3,  & 18.4940 & 26.4033 & 39.6142 & \cellcolor{lightgray} 0.99 \\
Binary search = true  & & & &\cellcolor{lightgray} \TBstrut\\ \hline

 C\&W, conf = 0 & 46.2162 & 0.7342 & 10.2935 & \cellcolor{lightgray} 0.96 \TBstrut\\ \hline

 C\&W, conf = 100 &  45.2334 & 0.8051 & 11.4638 & \cellcolor{lightgray} 0.97 \TBstrut\\ \hline

\end{tabular}}
\end{table}

\subsubsection{\textbf{Experimental Results on the $N_2$}}

In $N_{2}$ with test accuracy of 96.42\%, we notice that the adversary can successfully fool the network, even when considering a high number of convolutional layers. The CNNs are generally vulnerable to adversarial attacks when considered as TN. However, these networks have high classification accuracy. The experimental results illustrated in Table~\ref{attack_res_n2} demonstrate that most of the eight adversarial attacks have an ASR of more than 90\% and the transferability property is satisfied, i.e., the attacker can completely transfer the samples from the SN to the TN. Therefore, it is crucial to address this problem by providing suitable defense mechanisms that are efficient against well-known adversarial attacks.

\begin{table}[!h]
\centering
\footnotesize
\caption{Attack Results on the  $N_2$ when considered as TN. \label{attack_res_n2}}
\resizebox{1\columnwidth}{!}{
\begin{tabular}{|c|c|c|c|c|}
\hline
\textbf{Attack Type} & \textbf{PSNR} & \textbf{$L_1$ dist} & \textbf{Max. dist} & \cellcolor{lightgray} \textbf{ASR}   \TBstrut\\ \hline

 I-FGSM, $\varepsilon$ = 0.1 & 37.4549 & 2.6206 & 5.9534 & \cellcolor{lightgray} 1.00 \TBstrut\\ \hline

 FGSM, $\varepsilon$ = 0.1 &  28.1658 & 23.1140 & 41.0907 & \cellcolor{lightgray} 0.69 \TBstrut\\ \hline

 JSMA, $\theta$ = 0.01 &  54.2103 & 0.04465 & 13.3155 &  \cellcolor{lightgray} 0.96 \TBstrut\\ \hline

 BIM, $\varepsilon$ = 0.01 & 43.8285 & 1.8659 & 2.5848 & \cellcolor{lightgray} 0.99 \TBstrut\\ \hline

 LBFGS, $\varepsilon$ = 1e-5 & 61.0327 & 0.0760 & 3.0495 & \cellcolor{lightgray} 1.00 \TBstrut\\ \hline
 
 DeepFool, default parameter & 59.1931 & 0.1198 & 4.5238 &\cellcolor{lightgray}  0.58 \TBstrut\\ \hline

 PGD, $\varepsilon$ = 0.05, step size = 0.3, &  44.5092 & 2.0654 & 2.9510 & \cellcolor{lightgray} 0.98 \\
Binary search = true  & & & & \cellcolor{lightgray} \TBstrut\\ \hline

 C\&W, conf = 0 & 61.3460 & 0.0460 & 4.6834 & \cellcolor{lightgray} 0.96 \TBstrut\\ \hline

 C\&W, conf = 100 &  59.9532 & 0.0650 & 4.8708 & \cellcolor{lightgray} 1.00 \TBstrut\\ \hline

\end{tabular}}
\end{table}

%For each of the nine considered adversarial attacks; we computed the average PSNR, the average $L_{1}$ distortion, the average maximum absolute distortion, and the Attack Success Rate (ASR).
\section{Adversarial Defenses}
\label{Defenses}
In this section, we present and evaluate two different adversarial defense methods to mitigate adversarial attacks to improve TN's security: the MPAs approach and the FR approach. The MPA approach is one of the adversarial defense mechanisms that researchers recently considered to provide security for ML models. It consists of resisting the networks by fine-tuning the models~\cite{HigherOrder2017}. On the other hand, the FR approach aims to decrease the adversary's knowledge of the TN, and place the adversary in LK or SK by selecting various features for the classification task, i.e., selecting a random number of features from the whole feature space of the flattening layer. In what follows, we describe each of these methods.

\subsection{Most Powerful Attacks (MPAs)}

\begin{figure}[!h]
\begin{center}
\includegraphics[width=0.26\textwidth]{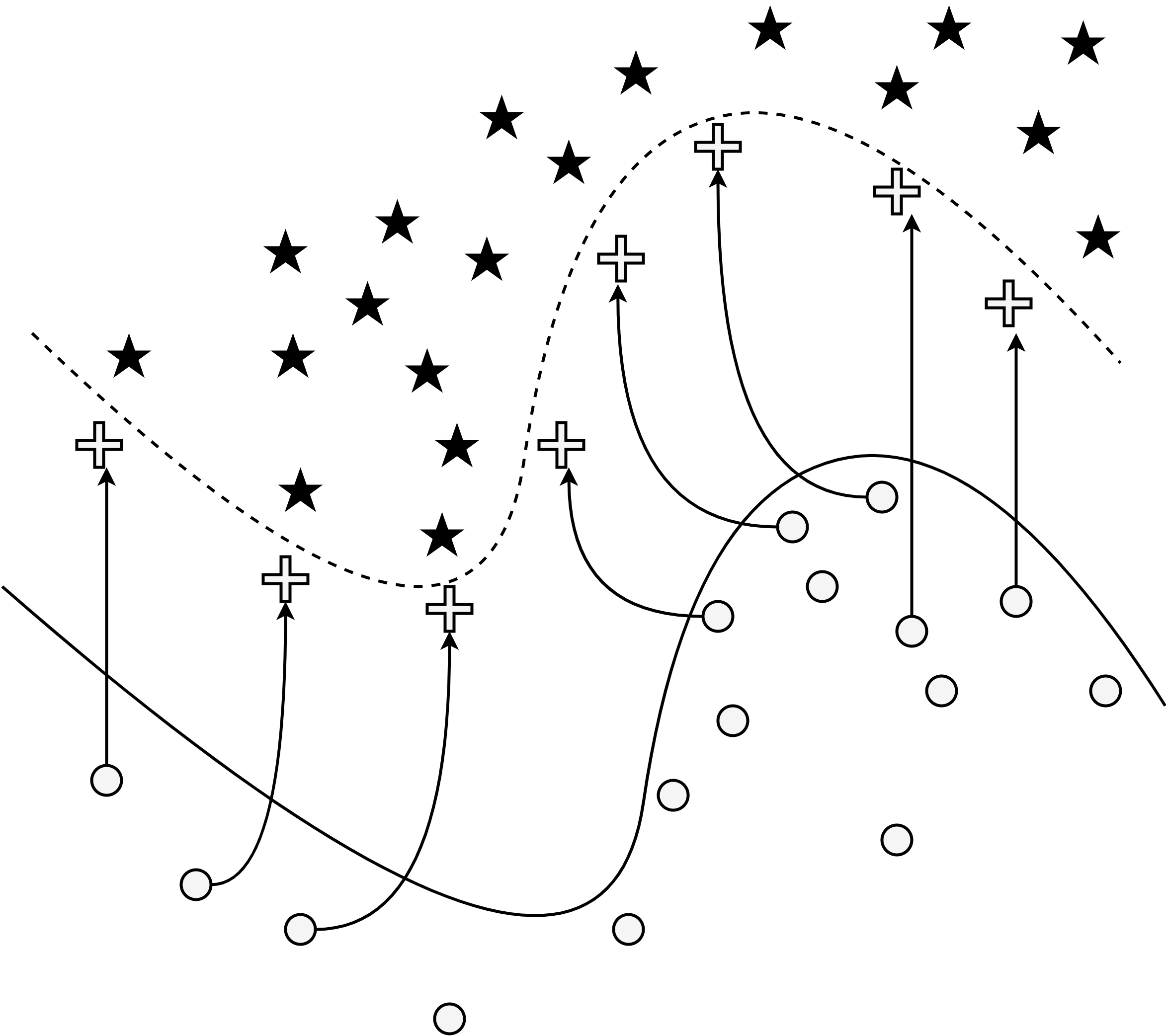}
\caption{
Representation of the MPA approach for an adversary-aware classifier. The generation of adversarial samples (crosses) enables decreasing the region of the benign samples (stars), thus challenging the obfuscation of dot samples as the star ones.}
\label{MPA}
\end{center}
\end{figure}

The rationale behind the MPA approach is to secure the TN by resisting it against the most powerful attacks (MPAs), i.e., the attacks that significantly reduce the model's accuracy. By leveraging MPA, with a high probability, we obtain a secure model against weaker attacks, thus, making the MPA approach efficient as it is not feasible to resist the detectors against all existing attacks \cite{HigherOrder2017}. The MPA approach consists of importing the attack samples into the training set that allows the decision margin to be refined. However, applying these new attacks to the detector is challenging. In~\cite{HigherOrder2017}, the authors proved that the choice of samples in the processing tools used for training is most effective in disabling the classifier's performance. As seen in Figure~\ref{MPA}, the orange samples are pristine, the green samples are manipulated, and the solid line illustrates the decision margin before fine-tuning. The red dots show the attacks that models were fine-tuned based on them. After fine-tuning, we observe that the decision margin completely changed (dotted line) and began closer to the pristine data; this new decision margin provides an extremely challenging situation for an attacker to cross the line as finding a gap between pristine data and decision margin is quite difficult. On the other hand, the experimental results in~\cite{Nowroozi2021AForensics} demonstrated that adding MPAs samples to the training set enables a good performance in the presence of a wider variety of attacks and processing. Although this strategy considers the SVM classifier and is already used for cybersecurity in Multimedia Forensics, we apply this approach in the context of computer networks and DL models. This is quite a novel use case of ML/DL models, which are usually considered due to their suitability and performance \cite{Demystifying}.

In~\cite{HigherOrder2017,Nowroozi2021AForensics}, the authors demonstrate most ML models, particularly deep learning networks, are inherently vulnerable and fragile against adversarial attacks. These vulnerabilities are critical in security-oriented applications given their negative impact on the performance of the models. In the MPA approach, we separately fine-tune $N_1$ and $N_2$ models with each attack sample. As we have eight types of attacks, we obtain eight new tuned models for each of $N_1$ and $N_2$. Then, we provide as input to each fine-tuned model the attack samples of other attacks to predict their labels (pristine or manipulated). Afterward, we gather and save the corresponding ASR. With this method, we can determine which fine-tuned model has more resistance against all the other attacks. For instance, consider the TN network $N_1$ fine-tuned by attack $A_1$ that is tested against attack $A_2$ and ASR is 10\%, it can be said that the fine-tuned model with $A_1$ can detect 90\% of $A_2$ attacks samples and can be fooled with 10\% of them.

\subsubsection{\textbf{Experimental Results for the MPA Approach}}

 We report the score results of the MPA approach for $N_{1}$ and $N_{2}$ models in Table~\ref{MPA_N1} and Table~\ref{MPA_N2}, respectively. Note that all the other scores of the MPA approach are equal to 1 except the reported results. For $N_{1}$, the tested results of the tuned adversarial attacks achieve a high score value for the attacks: I-FGSM010, FGSM010, BIM100, L-BFGS, JSMA001, JSMA, DeepFool, CW0, and CW100. Regarding the fine-tuned PGD005 attack, we remark that the test results have low scores for most of the tested attacks. However, for the deep network $N_{2}$, all the fine-tuned attacks have a high score value when tested against the eight adversarial attacks. According to the results, approximately all of the tuned models of the attacks were secure against other attacks; however, for verifying the MPA method, we need to apply the eight attacks on the fine-tuned models to verify if they can fool again the networks. 

\begin{table*}[!h]
\centering
\footnotesize
\caption{Score Results based on MPAs for $N_1$ model. \label{MPA_N1}}
\begin{tabular}{|c|c|c|c|c|c|c|c|c|c|c|}
\hline
$N_1$ MPAs Results & ---- & I-FGSM010 & FGSM010 & BIM100 & PGD005 & L-BFGS & JSMA001 & DeepFool & CW50 & CW100 \\ \hline
Tested with & BIM100 & 0.99 & 0.99 & 0.99 & 0.99 & 0.98 & 0.99 & 0.99 & 0.98 & 0.97 \\ \hline
 Tested with & PGD005 & 0.00 & 0.90 & 0.89 & 0.98 & 0.00 & 0.00 & 0.01 & 0.004 & 0.004 \\ \hline
\end{tabular}
\end{table*}

\begin{table}[!h]
\centering
%\footnotesize
\caption{Score Results based on MPAs for $N_2$ model \label{MPA_N2}}
\begin{tabular}{|c|c|c|c|}
\hline
$N_2$ MPAs Results & --- & I-FGSM010 & PGD005 \\ \hline
Tuned with & FGSM010 & 0.99 & 1 \\ \hline
- & BIM100 & 0.99 & 1 \\ \hline
- & JSMA001 & 0.95 & 0.98 \\ \hline
- & DeepFool & 0.99 & 1 \\ \hline
- & CW0 & 0.96 & 0.99 \\ \hline
- & CW100 & 0.96 & 0.98 \\ \hline
\end{tabular}
\end{table}

\subsubsection{\textbf{Security Level Evaluation for the MPA}}

After testing each of the tuned model with other adversarial attacks, we observe that all of the considered attacks achieve good results. However, by applying each attack again on the tuned networks, we remark that the adversary can perfectly fool the tuned networks. Therefore, we can confirm that the MPA approach is not efficient in securing $N_{1}$ and $N_{2}$ against adversarial attacks. Therefore, we consider in what follows another strategy based on the FR technique.

%%%%%%%%%%%%%%%%%%%%%%%%%%%%%%%%%%%%%%%%%%%%%%%%%%%%%%%%%%%%%%%%%%%%%%%%%%%%%%%%%%%%%%%%%%%%%%%%%%%%%%%%%%%%%%%%%%%%%%%%%%%%%%%%%%%%%%%%%%%%%%%%%%%%%%%%%%%%%
%%%%%%%%%%%%%%%%%%%%%%%%%%%%%%%%%%%%%%%%%%%%%%%%%%%%%%%%%%%%%%%%%%%%%%%%%%%%%%%%%%%%%%%%%%%%%%%%%%%%%%%%%%%%%%%%%%%%%%%%%%%%%%%%%%%%%%%%%%%%%%%%%%%%%%%%%%%%%

\subsection{Features Randomization (FR)}
In this approach, our goal is to provide a Limited-Knowledge condition for the attacker by selecting random features vectors of size ($F<N$), where F is the size of the random features vector and N is the size of the full features vector extracted from the SN flatten layer. Then, we give the TN the selected random features vector as input to perform the classification task (pristine or manipulate). This method satisfies the Limited-Knowledge setting as the attacker cannot guess the selected random features to perform adversarial attacks to fool the TN. Even when the whole feature set is examined ($F=N$), the detector differs from the classification architecture of the SN. In this case, as the SN and TN are similar, the attacker does not know the architecture of the TN. Moreover, the amount of attacker's information regarding the SN is smaller than in the MPA approach. To evaluate the FR approach with different random feature vectors, we consider different feature space sizes to identify which space size would provide higher security for the TN.

\subsubsection{\textbf{Experimental Results for the FR Approach}}

To implement the FR approach, we defined random feature space sizes by $F = \{5, 10, 30, 50, 200, 400, N\}$, where $N$ represents the flatten layer's full features size of $N_{1}$ and $N_{2}$, which are 1.728 and 3.200 respectively. In this study, we considered SVM as TN to train it with random feature vectors. For this purpose, we randomly selected 120.000 samples for training (60.000 for pristine and 60000 for manipulate), 10.000 samples for validation (5000 for pristine and 5.000 for manipulate), and 20.000 samples for testing (10.000 for pristine and 10.000 for manipulate). Given the computation cost challenges, we note that the SVM is not trained with all the 223.633 samples. Then, we fed the training, validation, and testing samples to SN and extracted full feature vectors of the samples from the flattened layer of the SN. Afterward, we selected random features vector from the full features vector 50 times for different sizes of $F$, i.e., at the end of this process, we had 50 different feature sets (training, validation, and test) for each $f \in F$. These feature sets are employed to train 50 SVMs for each $f \in F$. 

To better illustrate the FR approach, we assume $f=10$. To provide data for SVMs training, we consider 120.000 samples of training, 10.000 samples of validation, and 20.000 samples of the test that we feed to $N_{1}$. Then, we extract the features of its flattened layer (full features vector). Afterward, as the full features vector size is 1728 for $N_{1}$, we randomly select the features vector of size $f=10$ from $N=1728$ for each sample 50 different times, and which will be used to train 50 different SVMs. We repeat the same procedure for $N_{2}$.

When considering the SVMs as TN, we employed Radial Basis Function (RBF) kernel~\cite{rbfkernel} which has two hyper-parameters $C$ and $\gamma$ to control the error of classification and to give curvature weight of the decision boundary, respectively. As the hyper-parameters should be set before training the model, for finding the best $C$ and $\gamma$ for each SVM, we performed a 5-fold cross-validation via grid search. Then, we used the found hyper-parameters to train the SVM models. Next, we evaluated the SVMs with the test set features that did not include attacked samples. We ran our experiments on a computer with an Intel(R) Core i7 - 10 and 11 generation CPU with 32 GB of RAM. In addition, we performed the training and testing processes of all the SVMs using the LiBSVM library package~\cite{libsvm}. In Table.~\ref{average}, we report the average accuracies of 50 SVM models for each $f \in F$. 

%\begin{table}[!h]
%\centering
%\small
%\caption{Average percentage accuracy of SVMs in the absence of adversarial attacks.\label{average}}
%\begin{tabular}{l|lllllll|}
%\cline{2-8}
%\textbf{}                                           & \multicolumn{7}{c|}{\textbf{Size of Random Feature Vectors}}                                                                                                                                                                                                       \\ \cline{2-8} 
   %                                                 & \multicolumn{1}{c|}{\textbf{5}} & \multicolumn{1}{c|}{\textbf{10}} & \multicolumn{1}{c|}{\textbf{30}} & \multicolumn{1}{c|}{\textbf{50}} & \multicolumn{1}{c|}{\textbf{200}} & \multicolumn{1}{c|}{\textbf{400}} & \multicolumn{1}{c|}{\textbf{N}} \\ \hline
%\multicolumn{1}{|l|}{\textbf{\bm{$N_{1}$}}} & \multicolumn{1}{l|}{95.14}   & \multicolumn{1}{l|}{95.91}    & \multicolumn{1}{l|}{96.02}    & \multicolumn{1}{l|}{96.01}    & \multicolumn{1}{l|}{96.01}     & \multicolumn{1}{l|}{96.01}     & 96.01                        \\ \hline
%\multicolumn{1}{|l|}{\bm{$N_{2}$}}    & \multicolumn{1}{l|}{93.75}   & \multicolumn{1}{l|}{95.83}    & \multicolumn{1}{l|}{95.90}    & \multicolumn{1}{l|}{96.13}    & \multicolumn{1}{l|}{96.13}     & \multicolumn{1}{l|}{96.18}     & 96.18                        \\ \hline
%\end{tabular}
%\end{table}

%%
%%%
%%%%
\begin{table}[!h]
\centering
%\small
\caption{Average percentage accuracy of SVMs in the absence of adversarial attacks.\label{average}}
\begin{tabular}{|c|c|c|c|c|c|c|c|}
\hline
   ----    & 5 & 10 & 30 & 50 & 200 & 400 & N \\ \hline
$N_{1}$& 95.14 & 95.91 & 96.02 & 96.01 & 96.01 & 96.01 & 96.01 \\ \hline
$N_{2}$& 93.75 & 95.83 & 95.90 & 96.13 & 96.13 & 96.18 & 96.18 \\ \hline
\end{tabular}
\end{table}

To test the performance of the trained SVMs against attacked samples, we followed the same approach for getting the features from SN's flatten layer, i.e., we gave each attack sample to SN and extracted the features from the flattening layer. Then, we randomly selected 50 different feature sets for each features vector of size $f \in F = \{5, 10, 30, 50, 200, 400, N\}$. We considered 500 samples for each attack, including I-FGSM010, FGSM010, BIM100, L-BFGS, JSMA001, JSMA, DeepFool, C\&W0, and C\&W100. For example, we assume $f=30$ and select the FGSM010 attack. After applying the random feature selection procedure, we obtain 50 different feature vectors of size $500 \times 30$ for the FGSM010 attack that is ready to test the SVM models, i.e., to measure the ASR on TN.

To test the SVMs, we considered two methods with different knowledge levels for attackers: mis-match index testing and match index testing. In the mis-match index testing, the adversary has a Limited-Knowledge, while in the match index testing, the adversary has a Semi-Knowledge. In our study, we assume that the TN's model is secure if it can detect adversarial attacks with an accuracy of more than 60\%.

\noindent
\\
\textbf{Mis-match index testing:} In this procedure, the attacker has an LK condition due to the absence of knowledge regarding the TN model, the parameters of the TN model, and the random indices used for classifying. The attacker knows only the feature size, i.e., $f \in F$. Therefore, we tested each SVM model with the 50 randomly selected feature vectors of each attack. Then, we summed the score of these SVMs. As we have 50 SVMs, we calculated their average scores. In Algorithm~\ref{Algorithm1}, we present the mis-match index testing algorithm. In a mis-match index testing scenario, we considered $N_1$ as SN, the SVMs as TN, and we trained the SVMs by the randomly selected features vector. In Table~\ref{Shall_Mismatch_Results}, we provide the numerical results of the mis-match index testing. According to the reported results, it can be clearly shown that we achieved good results for the random feature size of 30 and 50. Moreover, in the full feature size $N=1728$, the adversary knows all the indexes, which is reasonable to obtain an insecure model. To that end, we state that using the random feature selection technique can increase the security model to some extent. 

\begin{algorithm}
\small
\label{Algorithm1}
\SetAlgoLined
\textbf{global} \textit{feature\_sizes} = [5, 10, 30, 50, 200, 400]\\
\textbf{global} \textit{attack\_names} = [IFGSM10, BIM100, PGD005, LBFGS, JSMA001, DeepFool, CW100, CW0, FGSM10]

\For{\textit{fs} \textbf{in} \textit{feature\_sizes}}{ 
 \For{\textit{attack} \textbf{in} attack\_names }{
        \textit{total\_mean} = 0\\
     \For{\textit{model\_num} \textbf{in} range(1,50) }{
 \textit{svm} = load\_svm\_model(\textit{model\_num})\\
        \textit{iter\_score} = 0\\
        \For{\textit{attack\_feature\_file} \textbf{in} range(1,50) }{
 \textit{file} = load\_attack(\textit{attack},\textit{attack\_feature})\\
    \textit{accuracy} = test\_svm(\textit{svm}, \textit{file})\\
    \textit{iter\_score} = \textit{accuracy} + \textit{iter\_score}
}
\textit{total\_mean} = \textit{total\_mean} +  (\textit{iter\_score} / 50 )
}
\textit{final\_mean} = \textit{total\_mean} / 50 \\
\texttt{/* The final mean is inserted in the table */}
}
}
 \caption{Mis-match index testing for Features Randomization}
\end{algorithm}

\begin{table}[!h]
%\scriptsize
\caption{Percentage of Mis-match index results for $N_1$ model (results reported in \%). \label{Shall_Mismatch_Results}}
\begin{tabular}{|c|c|c|>{\columncolor[gray]{0.8}}c|>{\columncolor[gray]{0.8}}c|c|c|c|}

\hline
 -- & 5 &  10 & 30 & 50 & 200 & 400 & N \\ \hline
 
 I-FGSM010 &51.07 & 58.59 & 76.23 & 75.08& 60.46 &72.09 & 0 \\ \hline
 
 BIM100 & 45.34 & 51.54 & 61.18  & 64.01 & 57.67 & 68.51 & 0 \\ \hline
 
 PGD005  & 45.43   & 51.43  & 62.01    & 65.37  & 58.5   & 69.29    & 0   \\ \hline
 
 L-BGFS   & 50.78   & 58.37 & 76.24 & 75.40   & 60.30   & 72.15  & 0          \\ \hline
 
 JSMA001  & 50.51     & 57.88      & 75.47     & 74.63     & 60.45      & 72.00    & 0       \\ \hline
 
 DeepFool & 49.56   & 56.74    & 72.54      & 72.66     & 60.14     & 71.42    & 0        \\ \hline
 
 CW100    & 50.82      & 58.34     & 75.98     & 75.18       & 60.31       & 72.11      & 0        \\ \hline
 
 CW0     & 50.76    & 58.31       & 76.05       & 75.29     & 60.31      & 72.11     & 0         \\ \hline
 
 FGSM010  & 51.07    & 58.59      & 76.23    & 75.08    & 60.46      & 72.09     & 0        \\ \hline
 
\end{tabular}
\end{table}

Similarly, we applied the mismatch index testing when considering $N_2$ as SN, and we report our numerical results in Table~\ref{Deep_Mismatch_Results}. The results show that we obtain good results for the random feature size of 200. Additionally, in the full future case, the adversary knows all the indexes, which is reasonable to have an insecure model. Therefore, by using the random feature selection, we can increase the security level of the TN to some extent.

\begin{table}[!h]
%\scriptsize
\caption{Percentage of Mis-match index results for $N_2$ model (results reported in \%). \label{Deep_Mismatch_Results}}
\begin{tabular}{|c|c|c|>{\columncolor[gray]{0.8}}c|>{\columncolor[gray]{0.8}}c|c|c|c|}
\hline
 
 -- & 5 & 10 & 30 & 50 & 200 & 400 & N \\ \hline
 
 I-FGSM010  & 11.60 & 19.43    & 32.53     & 37.19     & 70.03     & 54.08      & 0        \\ \hline
 
 BIM100  & 14.29   & 23.25     & 35.19    & 39.99     & 76.16     & 50.68       & 0.04    \\ \hline
 
 PGD005   & 14.86      & 23.83     & 35.44    & 40.03     & 75.31      & 56.37       & 0.05       \\ \hline
 
 L-BGFS    & 20.19    &30.54   & 37.46   & 41.72      & 66.51       & 44.84       & 0.21    \\ \hline
 
 JSMA001 & 27.35     & 37.41       & 43.99      & 47.99       & 66.88      & 49.85       & 0.58       \\ \hline
 
 DeepFool & 20.86    & 31.76   & 40.41     & 44.55      & 69.03 & 51.04     & 0.39      \\ \hline
 
 CW100    & 27.00    & 37.38 & 43.99     & 47.54  & 75.40     & 49.55     & 0.51     \\ \hline

 CW0      & 27.10     & 37.20     & 44.05     & 47.72      & 75.41       & 49.88      & 0.51      \\ \hline

 FGSM010  & 11.58     & 19.41     & 32.52      & 37.18       & 70.02        & 54.07       & 0          \\ \hline
 
\end{tabular}
\end{table}

\noindent
\\
\textbf{Match index testing:} In this scenario, the attacker has an SK condition. In other words, the TN classifier and its parameters are not available for the adversary. However, the information of selected indices and the features vectors size, i.e., $f \in F$, are accessible for the attacker. In this method, we tested each of the SVM models of feature size $f$ with one random features vector of each attack with the same index instead of 50 random features vectors. In what follows, we provide the Algorithm~\ref{Algorithm2} of match index testing. 

\begin{algorithm}
\small
\label{Algorithm2}
\SetAlgoLined
\textbf{global} \textit{feature\_sizes} = [5, 10, 30, 50, 200, 400]\\
\textbf{global} \textit{attack\_names} = [IFGSM10, BIM100, PGD005, LBFGS, JSMA001, DeepFool, CW100, CW0, FGSM10]

\For{\textit{fs} \textbf{in} \textit{feature\_sizes}}{ 
 \For{\textit{attack} \textbf{in} attack\_names }{
        \textit{total\_mean} = 0\\
     \For{\textit{model\_num} \textbf{in} range(1,50) }{
 \textit{svm} = load\_svm\_model(\textit{model\_num})\\
        
 \textit{file} = load\_attack\_feature(\textit{attack},\textit{model\_num})\\
    \textit{accuracy} = test\_svm(\textit{svm}, \textit{file})\\
\textit{total\_mean} = \textit{total\_mean} +   \textit{accuracy}
}
\textit{final\_mean} = \textit{total\_mean} / 50 \\
\texttt{/* The final mean is inserted in the table */}
}
}
 \caption{Match index testing for Features Randomization}
\end{algorithm}

In Table~\ref{Shall_Match_Results} and Table~\ref{Deep_Match_Results}, we report the numerical results in the match index testing of $N_1$ and $N_{2}$ when considered as SN models and SVM as TN, respectively. According to the presented results, we can claim that if the adversary has an SK on the features, it will be challenging to fool the TN. In fact, for the match index testing of the $N_{1}$, choosing the SVM between 200 and 400 will provide a security level of the TN by more than 60\% for all the considered adversarial attacks. For $N_{2}$, we notice that the SVM of size 200 enables a good security level for TN with more than 61\% for all the adversarial attacks.

%
%%
%%%%
\begin{table}[!h]
\scriptsize
\caption{Percentage of Match index results for $N_1$ model (results reported in \%). \label{Shall_Match_Results}}
\begin{tabular}{|c|c|c|>{\columncolor[gray]{0.8}}c|>{\columncolor[gray]{0.8}}c|c|c|c|}
\hline

  --& 5 & 10 & 30 & 50 & 200 & 400 & N \\ \hline
  I-FGSM010 & 66.10      & 58.02     & 73.30     & 71.58    & 60.08      & 69.12       & 0.00     \\ \hline
BIM100   & 46.62    & 53.14     & 63.57      & 56.87     & 62.53       & 70.36       & 0.04      \\ \hline
PGD005  & 45.36    & 53.34       & 63.96      & 57.96     & 63.66       & 70.71      & 0.05       \\ \hline
L-BGFS    & 77.51     & 58.91      & 72.22     & 71.70     & 60.42      & 69.86       & 0.21      \\ \hline
JSMA001 & 71.20    & 55.17      & 73.98      & 71.79     & 60.64      & 70.24      & 0.58      \\ \hline
DeepFool& 69.65     & 56.86     & 67.55      & 71.86      & 59.77       & 68.64       & 0.39     \\ \hline
CW100   & 75.34    & 57.48      & 72.30      & 71.70      & 60.44        & 69.77       & 0.51      \\ \hline
CW0     & 76.78     & 57.85      & 72.38     & 71.56      & 60.55      & 69.72       & 0.51     \\ \hline
FGSM010  & 66.10     & 58.01     & 73.29     & 71.58      & 60.08      & 69.12        & 0.00    \\ \hline

\end{tabular}
\end{table}

\begin{table}[!h]
\scriptsize
\caption{Percentage of Match index results for $N_2$ model (results reported in \%). \label{Deep_Match_Results}}
\begin{tabular}{|c|c|c|>{\columncolor[gray]{0.8}}c|>{\columncolor[gray]{0.8}}c|c|c|c|}
\hline
 --& 5 & 10 & 30 & 50 & 200 & 400 & N \\ \hline
 
 I-FGSM010 & 15.51 & 22.97  & 40.86  & 38.54 & 63.43 & 53.35       & 0.00      \\ \hline
BIM100 & 22.62    & 33.05       & 79.61    & 49.67   & 66.87       & 57.67        &0.04      \\ \hline
PGD005  & 23.70      & 34.80      & 50.31      & 50.20      & 67.51      & 57.80       & 0.05      \\ \hline
L-BGFS  & 34.73     & 48.77      & 50.75      & 46.88       & 64.98        & 33.18       & 0.21      \\ \hline
JSMA001 & 79.20     & 83.99     & 77.66     & 70.33      & 70.96      & 57.88      & 0.58      \\ \hline
DeepFool & 43.01     & 56.28      & 67.07    & 62.46       & 61.73      & 48.83       & 0.39      \\ \hline
CW100   & 79.75     & 85.50     & 77.18      & 71.84      & 70.30      & 47.79       & 0.51      \\ \hline
CW0      & 80.30      & 84.97     & 77.29      & 72.92      & 70.75      & 47.97      & 0.51      \\ \hline
FGSM010  & 75.47     & 22.93      &40.83    & 38.52      & 67.39      & 53.34      & 0.00      \\ \hline

\end{tabular}
\end{table}

\subsubsection{\textbf{Security Level Evaluation for the FR approach}}

According to the reported results, we observe that the mis-match index case is more robust than the match index case. This could be explained by the fact that each model was tested with 50 files, which is more likely to be secure against different attacks feature. Therefore, the FR approach's experimental results demonstrate that the TN security level is improved compared to the MPAs approach.

\section{Discussion}
\label{Discussion}
In this work, we applied two different defense strategies against eight adversarial attacks with different parameters under black-box settings, namely: The JSMA, the PGD attack, the L-BFGS attack, the I-FGSM attack, the FGSM attack, the DeepFool attack, the BIM attack, and the C\&W attack (with the strength parameters 0 and 100).
The first strategy is based on the MPA (i.e., the source and target network architectures are similar), while the second defense mechanism is based on the features randomization technique (i.e., the source and target network architectures are different). 

Given that prior works demonstrated the robustness of the MPA approach against adversarial manipulations, notably in the category of the adversary-aware detector. We imported the MPA approach to our study to evaluate its performance using the fine-tuning technique in the field of network security. However, we found that such an approach is inefficient, and the TN model cannot resist the adversarial transferability. In this context, we proved that all the eight considered adversarial attacks were transferred from the SN to the TN.

To that end, we developed a novel strategy based on the feature randomization technique. In this case, we investigated the potential of utilizing FR to increase the resilience of DL models against adversarial cases by limiting the attack transferability. We applied our approach in a wide range of scenarios, demonstrating that the FR approach can significantly reduce the transferability of adversarial attacks, thereby enhancing the security of the DL models. Even though, in certain circumstances, we found that the mis-match in structure between the SN and the TN is sufficient to prevent adversarial transferability. Additionally, our investigations demonstrated that for the small size of random feature vectors, the complexity increases in the training data for the TN. This complexity could be explained by the high amount of random cases, which can decrease the adversary's awareness of training data. Therefore, the TN can resist more adversarial attacks and prevent attack transferability. Interestingly, we decreased the attacker's knowledge in the FR approach by changing the TN architecture and employing a TN model different from the SN model. Consequently, the attacker has limited awareness of the TN model and its parameters. 
For instance, we considered the SVM as TN, which had different architecture from SN models ($N_1$ and $N_2$). 
In this case, we decreased the adversary's knowledge of the LK condition. Therefore, the attacker must perform deep searches to determine TN's architecture and parameters. Moreover, we showed that the FR approach is efficient in the SK scenario, i.e., the attacker has information about the randomly selected indices for training the TN and their feature vector size. Accordingly, we analyzed the SK scenario through match index testing. Our numerical results demonstrated that the DL models are secure against attack transferability issues when the adversary is under SK condition. Furthermore, we evaluated the computational cost of the attacks for 500 samples with the trained networks. As shown in Table~\ref{cost1}, we remark that the computational cost of the C\&W attack is significantly higher than other adversarial attacks in the SN and TN. Further, we also evaluated the runtime for training 50 SVM based on the features extracted from $N_1$ and $N_2$ in the FR approach. As depicted in Table~\ref{cost2}, the results show that the computational cost increases when the number of features increases

%%
%%%%
%%%%%%%
\begin{table}[!h]
\centering
\normalsize
\caption{Computational cost of the Adversarial attacks for 500 samples on $N_1$ and $N_2$ in seconds}\label{cost1}
\begin{tabular}{|c|c|c|}
\hline
Attack Type & $N_1$ & $N_2$ \\ \hline
I-FGSM010 & 1500 & 500 \\ \hline
FGSM010 & 500 & 500 \\ \hline
JSMA001 & 5000 & 3600 \\ \hline
BIM100 & 500 & 500 \\ \hline
L-BFGS & 9000 & 6000 \\ \hline
 DeepFool& 500 & 500 \\ \hline
PGD005 & 3000 & 1800 \\ \hline
CW0 & 59000 & 45000 \\ \hline
CW100 & 60000&47000   \\ \hline
\end{tabular}
\end{table}

%
%%%
%%%%%
\begin{table}[!h]
\centering
%\normalsize
\caption{Computational cost in seconds for training 50 SVM based on the features extracted from $N_1$ and $N_2$}\label{cost2}
\begin{tabular}{|c|c|c|}
\hline
Number of Features & $N_1$ & $N_2$ \\ \hline
5 & 8100 & 8100 \\ \hline
10 & 18000 & 18000 \\ \hline
30 & 45000 & 45000 \\ \hline
50 & 48000 & 48000 \\ \hline
200 & 189000 & 189000 \\ \hline
400 & 468000 & 468000 \\ \hline
Full features for one SVM & 22800 & 12600 \\ \hline
\end{tabular}
\end{table}

\section{Conclusion and Future Work}
\label{Conclusion}
Over the past decades, the increasing applications of machine and deep learning have triggered the need to consider its robustness against adversarial attacks. In this context, the adversary can craft malicious samples and transfer the adversarial attacks from the SN to the TN. To avoid the transferability property, it is crucial to improve the target network's security. In this paper, we investigated evasion attacks on ML/DL models applied in the testing phase. We leveraged the potential of utilizing the feature randomization technique to increase the resilience of DL models against adversarial samples, thus impeding attack transferability. Our experimental results in LK and SK conditions demonstrated that the FR approach could significantly reduce the transferability of adversarial attacks, thereby protecting the TNs from adversarial manipulations. We also showed that in some cases, the architectural difference between the SN and the TN is satisfactory to avoid adversarial transferability. Our future work will focus on poisoning attacks and their transferability. In particular, where an attacker may poison the training data by inserting precisely selected samples, ultimately threatening the entire learning process. The poisoning process can thus be viewed as malicious contamination of the training data. Further, we aim to design a defense mechanism against backdoors attacks against ML/DL models. These attacks hunt for high correlations in the training data without checking for causative variables.

\bibliographystyle{IEEEtran}

\bibliography{References}
\begin{IEEEbiography}
[{\includegraphics[width=1in,height=1.25in]{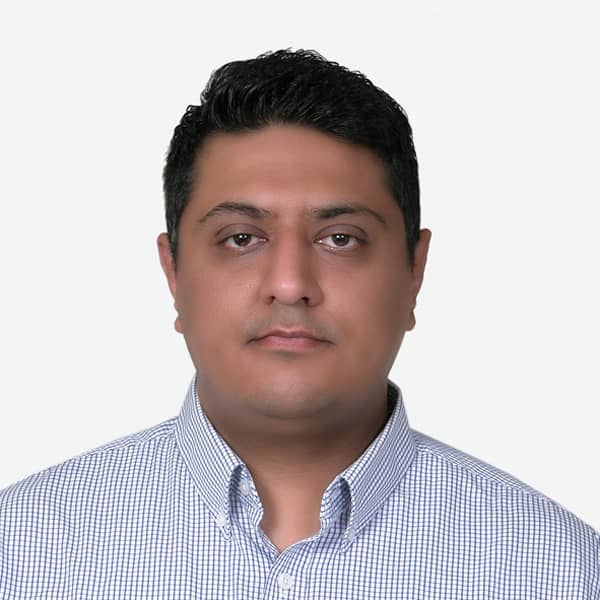}}]{Ehsan Nowroozi} is an Assistant Professor at Istanbul's Bahçeşehir University's Faculty of Engineering and Natural Sciences, Department of Computer Engineering. He is also affiliated with Padua and Siena Universities in Italy, and he collaborates as an external researcher with SPRITZ and the VIPP research group. In 2020, he received his Ph.D. from the University of Siena in Italy, and he successfully defended his Ph.D. Following his Ph.D., he worked as a postdoctoral fellow at Siena University's VIPP research group until 2021, then at Padua University's SPRITZ research group until 2022, and finally at Sabanci University in Turkey as a postdoctoral fellow until 2022. His primary research interests are in the areas of security and privacy, namely the use of image processing techniques to multimedia authentication (multimedia forensics), network and internet security, and so on. His professional service and activity as a reviewer for IEEE TNSM, IEEE TIFS, IEEE TNNLS, EURASIP Journal on Information Security, and other journals. He is also a member of the IEEE Signal Processing Society and the IEEE Young Professionals.%received the Ph.D. degree from Siena University, Italy. He is a a Postdoctoral researcher with the Faculty of Engineering and Natural Sciences (FENS), Center of Excellence in Data Analytics (VERIM), Sabanci University, Istanbul, Turkey 34956. He was a Postdoctoral Researcher with Siena (VIPP Group) and Padova University (SPRITZ Group), Italy 2020 and 2021, respectively. His Ph.D. and a Postdoctoral researcher at Siena University were funded by DARPA, and a Postdoctoral researcher at Padova University was founded by the STAR program. His main research interest is in the area of Security and Privacy with particular reference to the application of image processing techniques to authentication of multimedia (multimedia forensics), network and internet security, etc. His professional service and activity as a reviewer of the journal IEEE TNSM, IEEE TIFS, IEEE TNNLS,  Elsevier journal Digital, EURASIP Journal on Information Security, etc. He is also a member of the IEEE Young Professionals and IEEE Signal Processing Society.
\end{IEEEbiography}

\begin{IEEEbiography}[{\includegraphics[width=1in,height=1.25in]{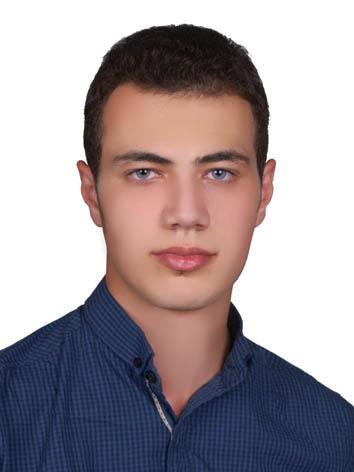}}]{Mohammadreza Mohammadi}  is a second-year Master student of ICT at University of Padua. He received the bachelor’s degree in computer engineering(software-network) from Bu-Ali Sina University, Hamedan, Iran in 2019. His main research interest is in the area of Machine Learning, Cybersecurity, IoT and Computer Vision. In addition, he works on reserach contexts which are combination of Industrial IoT security and artificial intelligence, and Intrusion detection systems(IDS) and Healthcare systems and He is also a graduate student member of IEEE institution.
\end{IEEEbiography}

\vskip -1\baselineskip plus -1fil
\begin{IEEEbiography}[{\includegraphics[width=1in,height=1.25in]{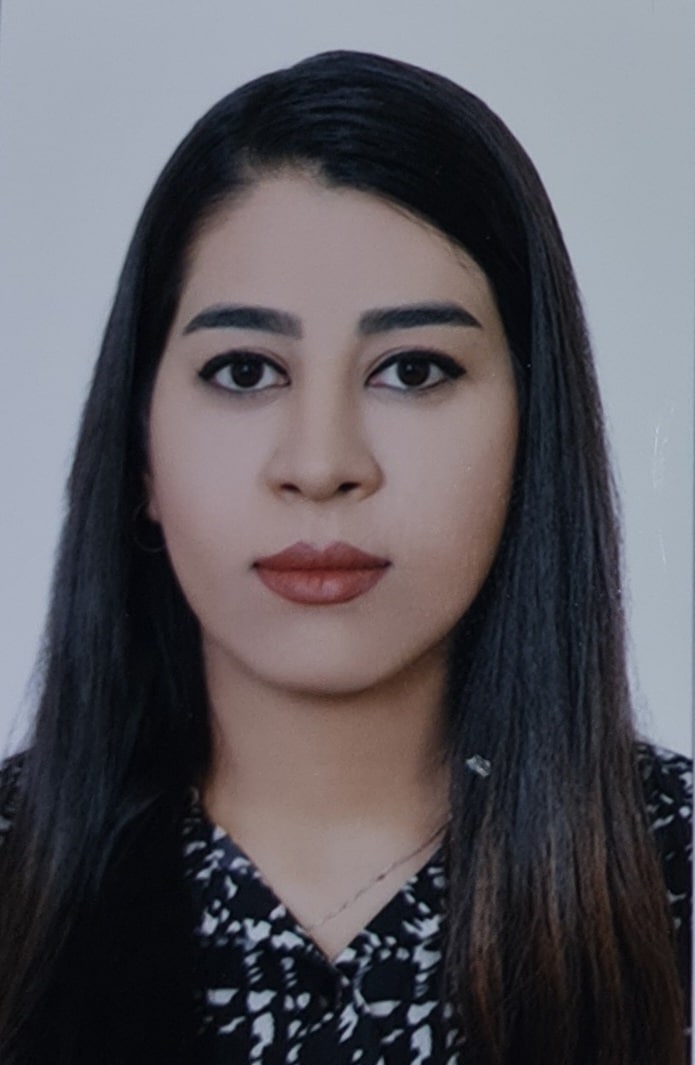}}]{Pargol Golmohammadi} is a graduate student member of IEEE institution. She received the bachelor’s degree in biomedical engineering(bioelectric) from Hamedan University of Technology, Hamedan, Iran in 2018. Now, She is a second-year Master student of ICT(life and health) at University of Padua. Her main research interest is in the area of Biometrics, IoT, E-Health Security and Machine Learning.  
\end{IEEEbiography}

\vskip -2\baselineskip plus -1fil

\begin{IEEEbiography}[{\includegraphics[width=1in,height=1.25in]{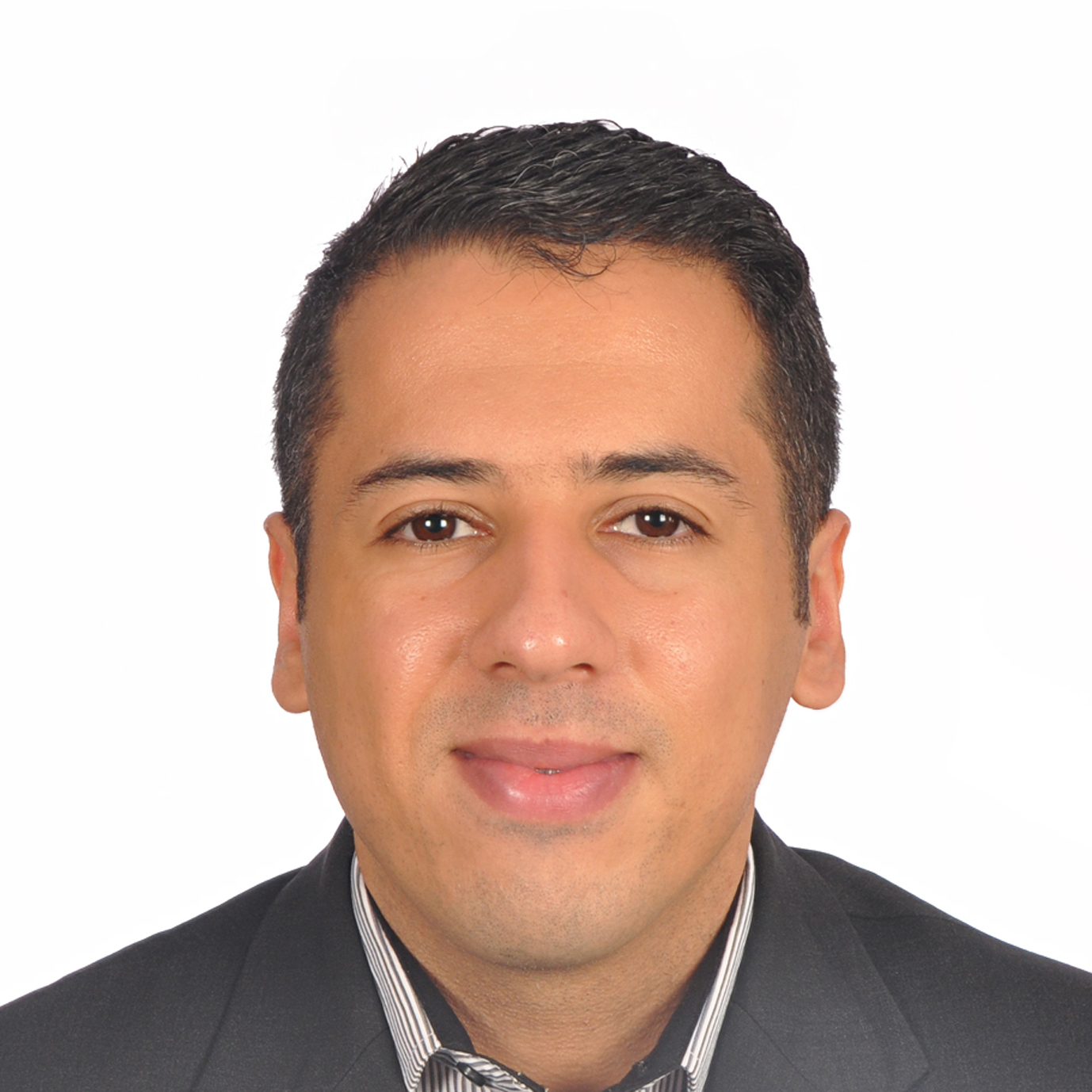}}]{Yassine Mekdad} received a Msc Degree in Cryptography and Information Security from Mohammed V University of Rabat, Morocco. He holds a guest researcher position with the SPRITZ research group at University of Padua, Italy. He is currently working as a Research Scholar at the Cyber-Physical Systems Security Lab (CSL) at Florida International University, Miami, FL, USA. His research interest principally cover security and privacy problems in the Internet of Things (IoT), Industrial Internet-of-Things (IIoT), and Cyber-physical systems (CPS). Furthermore, he works on research problems at the intersection of the cybersecurity and networking fields with an emphasis on their practical and applied aspects. 
\end{IEEEbiography}
\vskip -2\baselineskip plus -1fil
\begin{IEEEbiography}[{\includegraphics[width=1in,height=1.25in]{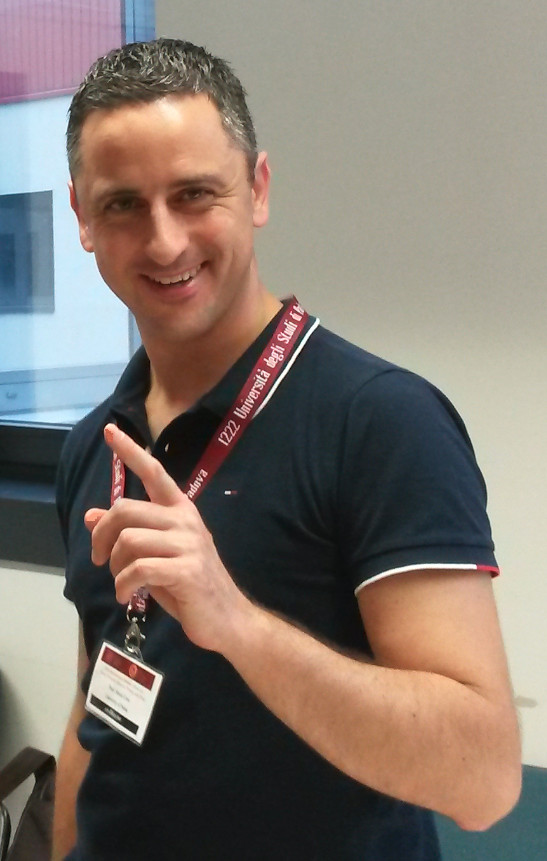}}]{Mauro Conti} is Full Professor at the University of Padua, Italy. He is also affiliated with TU Delft and University of Washington, Seattle. He obtained his Ph.D. from Sapienza University of Rome, Italy, in 2009.
After his Ph.D., he was a Post-Doc Researcher at Vrije Universiteit
Amsterdam, The Netherlands. In 2011 he joined as Assistant Professor at
the University of Padua, where he became Associate Professor in 2015,
and Full Professor in 2018. He has been Visiting Researcher at GMU,
UCLA, UCI, TU Darmstadt, UF, and FIU. He has been awarded with a Marie
Curie Fellowship (2012) by the European Commission, and with a
Fellowship by the German DAAD (2013). His research is also funded by
companies, including Cisco, Intel, and Huawei. His main research
interest is in the area of Security and Privacy. In this area, he
published more than 450 papers in topmost international peer-reviewed
journals and conferences. He is Editor-in-Chief for IEEE Transactions on
Information Forensics and Security, Area Editor-in-Chief for IEEE
Communications Surveys \& Tutorials, and has been Associate Editor for
several journals, including IEEE Communications Surveys \& Tutorials,
IEEE Transactions on Dependable and Secure Computing, IEEE Transactions
on Information Forensics and Security, and IEEE Transactions on Network
and Service Management. He was Program Chair for TRUST 2015, ICISS 2016,
WiSec 2017, ACNS 2020, CANS 2021, and General Chair for SecureComm 2012,
SACMAT 2013, NSS 2021 and ACNS 2022. He is Fellow of the IEEE, Senior
Member of the ACM, and Fellow of the Young Academy of Europe.
\end{IEEEbiography}
\vskip -2\baselineskip plus -1fil
\begin{IEEEbiography}[{\includegraphics[width=1in,height=1.25in]{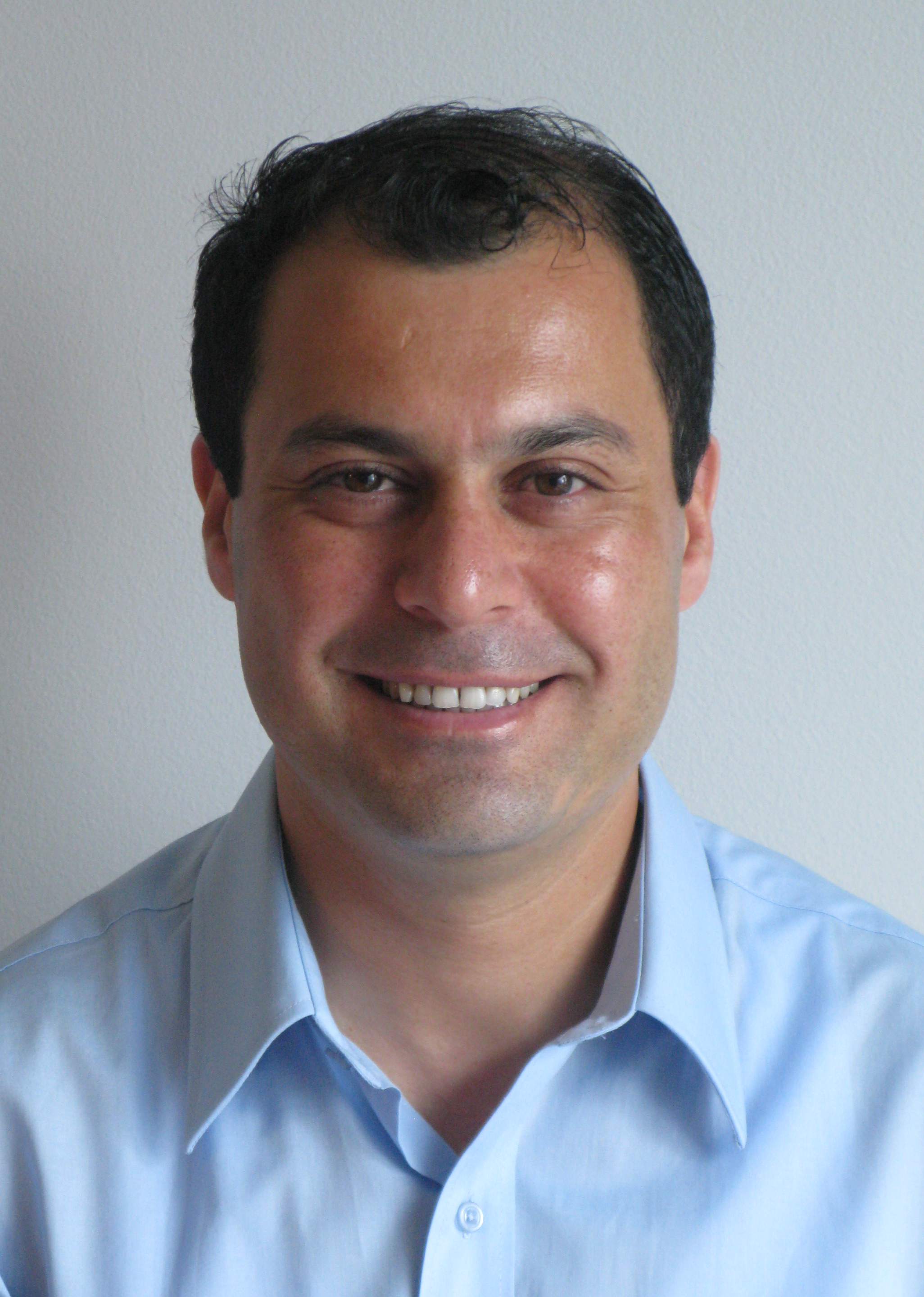}}]{A. Selcuk Uluagac} is an Eminent Scholar
Chaired Associate Professor and the director of Cyber-Physical Systems Security Lab in the Department of Electrical and Computer Engineering at Florida International University, Miami,
Florida, USA. Before FIU, he was a Senior Research Engineer at Georgia Institute of Technology and at Symantec. He holds a M.S. and Ph.D.
from Georgia Tech and an M.S. from Carnegie Mellon University. He is expert on security and privacy topics with hundreds of scientific/creative
works in practical and applied aspects of these areas. He received US NSF CAREER Award (2015), US Air Force Office of Sponsored Research's Summer Faculty Fellowship (2015), and University of Padova's
(Italy) Summer Faculty Fellowship (2016). His research in cybersecurity has been funded by numerous government agencies and industry. He has served on the program committees of top-tier security conferences such as IEEE Security \& Privacy ("Oakland"), NDSS, Usenix Security, inter alia. He was the General Chair of ACM Conference on Security and Privacy in Wireless and Mobile Networks (ACM WiSec) in 2019. Currently, he serves on the editorial boards of IEEE Transactions on Mobile Computing, Elsevier Computer Networks Journal, and the IEEE Communications and Surveys and Tutorials (network security lead).
\end{IEEEbiography}
\vskip -2\baselineskip plus -1fil
\begin{comment}

\end{comment}
%\vskip -2\baselineskip plus -1fil

%%%%%%%%%%%%%%%%%%%%%%%%%%%%%%%%%%%%%%%%%%%%%%%%%%%%%%%%%%%%%%%%%%%%%%%%%%%%%%%%
\end{document}